# Low-temperature fabrication of amorphous carbon films as a universal template for remote epitaxy

T. Henksmeier[1,2*], P. Mahler[1], A.Wolff[1], D. Deutsch[1], M. Voigt[3], L. Ruhm[3], A. M. Sanchez[4], D. J. As[1], G. Grundmeier[3], D. Reuter[1,2]

[1]*Paderborn University, Department of Physics and Center for Optoelectronics and Photonics Paderborn (CeOPP), Warburger Str. 100, 33089 Paderborn, Germany*
[2]*Institute for Photonic Quantum Systems (PhoQS), Paderborn University, 33098 Paderborn, Germany*
[3]*Paderborn University, Department of Chemistry, Warburger Str. 100, 33089 Paderborn, Germany*
[4]*Warwick University, Department of Physics, Coventry CV4 7AL, UK*

*Corresponding author: tobias.henksmeier@upb.de

**ABSTRACT**

We report on the low-temperature fabrication (300 °C) of ultrathin 2D amorphous carbon layers on III-V semiconductors by plasma-enhanced chemical vapor deposition as a universal template for remote epitaxy. We present growth and detailed characterization of 2D amorphous carbon layers on various host substrates and their subsequent remote epitaxial overgrowth by solid-source molecular beam epitaxy. We present the fabrication of ultra-smooth monolayer thick amorphous carbon layers with roughness ≤0.3 nm determined by atomic-force microscopy and X-ray reflectivity measurement. We show that precisely tailoring the carbon layer thickness allows superior tunability of the substrate-layer interaction. Further, X-ray photoelectron and Raman spectroscopy measurements reveal predominantly $sp^2$-hybridised carbon in the amorphous layers. We observe that a low-temperature nucleation step is favorable for nucleation of III-V material growth on substrates coated with thin amorphous carbon layers. Under optimized preparation conditions, we obtain high-quality, single-crystalline, (001)-oriented GaAs, cubic-AlN, cubic-GaN and $In_xGa_{1-x}As$, respectively, and various carbon-coated (001) oriented substrates as GaAs, InP and 3C-SiC. Transmission electron microscopy images of the substrate-carbon-layer interface reveal a stretching of the atomic bonds at the interface and high-resolution X-ray diffraction measurements reveal high crystal quality and low dislocation densities $<1 \times 10^7$ cm$^{-2}$. Our results show the universality of our carbon deposition process to fabricate templates for remote epitaxy, e. g., for remote epitaxy on temperature sensitive substrates like GaAs or InP and growth of metastable phases. Lift-off of layers from their substrates is demonstrated by employing a Ni stressor.

**Introduction**

Epitaxy of a crystalline layer on a crystalline substrate is widely employed to produce semiconductor heterostructures and is extremely important nowadays for the production of functional semiconductor devices. However, the thermal and lattice mismatch of substrate and epitaxial layer often limits the fabrication of more variable devices. Further, the fabrication of epitaxial devices is challenging and expensive despite several techniques to detach any epitaxial layer from an expensive substrate allowing subsequent substrate refurbishment [1]. Van-der-Waals epitaxy and remote epitaxy were recently proposed as way to weaken the substrate-layer binding allowing to peel-off the epitaxial film after growth [2-6]. In van-der-Waals epitaxy substrate and layer are completely decoupled avoiding/reducing the formation of dislocations but often the crystal quality suffers from poor surface wettability resulting in the formation of polycrystalline films for instance for growth of group-III arsenide semiconductors [7]. Remote epitaxy has been presented as an approach to partially weaken the substrate-layer-interaction but still transferring the crystal orientation to the growing layer [6, 9] so a film with high crystalline quality can be realized. To date, remote epitaxy on 2D materials has been extensively explored for the fabrication of freestanding semiconductor membranes and heterogeneously integrated devices [3,5,6,8-10]. High-quality film growth and fabrication of functional heterostructures has been demonstrated in several semiconductor material systems e.g. group-III arsenides [3,6,8,11], group-III nitrides [8,11-14] and oxides [15-17]. The growing layer is separated by a thin 2D material like graphene or h-BN from the substrate allowing to peel-off layers from different substrates for monolithic 3D integration fabrication [19, 20]. The main advantage of remote epitaxy is the substrate re-usage after the epitaxial layer lift-off, therefore reducing production cost of freestanding heterostructures and allowing fabrication of versatile structures [3]. Mismatched film growth on monolayer graphene was also demonstrated. Besides strain relaxation by the formation of dislocation a layer slip over the graphene was proposed as a relaxation pathway of mismatched layers [18]. However, layers exhibit still significant dislocation densities. While for layer growth on SiC a monolayer graphene can be formed by precisely tuning the sublimation conditions [21, 22], for other substrates epitaxial or CVD graphene has to be transferred on the substrate which is difficult to realize on wafer-scale and further comes along with its own challenges like pinholes and wrinkles in the 2D material thus leading to pinhole-seeded epitaxy or impede high quality remote epitaxy [23,24]. Based on the general understanding of remote epitaxy derived with 2D materials, remote epitaxy was demonstrated on transfer-free, direct grown amorphous carbon (a-C) layers [3]. Thin a-C layers were deposited on $Al_{0.5}Ga_{0.5}As$ in MOCVD employing toluene and thin boron nitride (BN) films were deposited on GaN in MBE employing elemental boron and nitrogen plasma thus providing a cheap and scalable process to produce templates for remote epitaxial growth [3]. The key for remote epitaxy is the dominant $sp^2$-hybridized in-plane bonding in the a-C and BN layer comparable to that of the 2D material counterparts. However, subsequent overgrowth of these amorphous layers lead to larger defect densities in the epitaxial film. Not optimized nucleation parameters and thickness control of the amorphous layers were discussed as possible reasons [3]. Further, the employed high temperatures required in the process reported in [3] for a-C or BN growth are not compatible with temperature sensitive substrates, e. g., InP which is an important substrate for the fabrication of devices operating in the infrared spectral range. A superior method for the fabrication of 2D materials on arbitrary host substrates as a universal remote epitaxial template should include a low-temperature, low-cost, high throughput, and complete vacuum compatibility 2D-material deposition process [10]. Overgrowth of the produced 2D material layer must enable high crystal quality growth.

In this work, we present a universal method for the remote epitaxial growth of semiconductor membranes based on $sp^2$-hybridised a-C layers and demonstrate the remote heteroepitaxial overgrowth with III-As and cubic III-N films. First, we present the fabrication of thin a-C layers on various substrates by plasma-enhanced chemical vapor deposition (PECVD) following subsequent annealing at rather low-temperature (300 °C) in UHV demonstrating a universal, cheap and fast way for the fabrication of high-quality remote epitaxy templates on arbitrary substrates, e. g., temperature sensitive GaAs and InP. We present detailed characterization of the a-C layer and show a route for optimized a-C layer fabrication following subsequent overgrowth by group-III arsenides and cubic group-III nitrides layers on a-C covered GaAs(001), InP(001) and 3C-SiC/Si(001) substrates, respectively. X-ray reflectivity (XRR), atomic force microscopy (AFM) and Raman spectroscopy measurements were used to precisely characterize the a-C layer thickness, the carbon binding state and the surface roughness. We demonstrate tunability of the substrate-layer interaction by tailoring the a-C layer thickness and show that besides the substrate ionicity, the growth temperature and the ad-atom species on the surface guides the remote epitaxial nucleation on a-C covered substrates. In the second part of this contribution, we show the wide application range of our a-C layer preparation method by demonstrating growth of high-quality $In_xGa_{1-x}As$ layers on a-C covered InP substrates and the growth of metastable group-III nitrides on a-C covered 3C-SiC/Si(001) quasi-substrates. Finally, we discuss the strain relaxation on a-C exemplary for highly mismatched $In_{0.5}Ga_{0.5}As$ layers grown on a-C covered substrates.

**Fabrication of a-C layers**

To fabricate the a-C layers, we employed a PECVD process (see Methods) employing a methane ($CH_4$)-argon (Ar) gas mixture and a subsequent UHV annealing step (300 °C) in our MBE system. Our a-C deposition process is low-cost, scalable and avoids surface contamination as it is performed in vacuum. It is suitable for temperature sensitive substrates like group-III arsenides (i.e. InP, GaAs) as the a-C deposition is done at room-temperature following subsequent annealing in UHV at 300°C. The PECVD deposition process can be integrated in a cluster with other systems operated in vacuum like MBE growth chambers to avoid any air exposure of the a-C layer. The $CH_4$ molecules are cracked by using a large ICP-power while a small RF-power is added to direct the ions to the sample surface, thus keeping the sample ion bombardment damage low. Carbon deposition with 0 W-10 W RF-power was tested. To achieve similar a-C layer thicknesses for smaller RF-powers, the deposition time had to be adjusted while all other parameters were kept the same. We measured similar surface roughness and Raman-spectra for all a-C layers independent of the RF power used. Fig. 1 a) shows the XRR-measurement of entire ¼ of 3" wafers of a-C covered GaAs substrates fabricated with different carbon deposition times together with the corresponding fitted profiles of the observed Kiessig fringes. We tested to fit a simple two-layer model consisting of an a-C layer on a GaAs substrate and a more advanced model of two a-C layer of different density and with a <1 nm thin a-C layer directly on the GaAs substrate. The latter model resulted in smaller residuals; fits and measured data agree well allowing precise determination of the total a-C layer thicknesses. Fig. 1 b) shows that the overall a-C layer thickness increases linearly with the deposition time and thus is a not self-limiting process, which allows to tailor the a-C layer thickness but also requires precise optimization to reach monolayer-like a-C layer thickness. We performed XRR measurements at different positions across the ¼ 3" GaAs wafers (see Supplementary Note 1 and Supplementary Fig. 1 a)) to identify the a-C thickness homogeneity across the substrate. The a-C thickness deviation is <1 % around the wafer centre (roughly 20x20 mm); this is the part of the wafer we analysed with all other methods presented in this manuscript. However, we also observed, that the a-C thickness decreases by roughly 10% toward the wafer edges which is due to the PECVD table size limitation and thus slightly different plasma conditions toward the substrate edge. We further confirmed the deposition of a-C on GaAs by AFM measurements employing a shadow mask approach discussed in the Supplementary Note 1 and Supplementary Fig. 1 b)-d). The surface topology of the fabricated a-C layers was measured by AFM revealing a smooth surface with a root-mean-square (RMS)-roughness of ≤0.3 nm equal to the surface roughness of ≤0.3 nm of our bare GaAs substrates (see Supplementary Fig. 2 a)) proving homogenous and smooth carbon deposition. To mimic the substrate-layer interaction gap introduced by a monolayer graphene of roughly 0.5 nm [6], we expect an a-C layer thickness of approximately 0.5 nm as suitable to provide a template for high-quality remote-epitaxy. This corresponds to deposition times <30 sec for 10 W RF-power regarding the data shown in Fig. 1 b). To precisely distinguish between a thick a-C layer, a quasi-monolayer and partial a-C coverage of the GaAs substrates, we exploit the thermal oxide evaporation of GaAs substrates at elevated temperatures in UHV: We first deposited a-C on a GaAs substrate, then exposed the a-C covered GaAs sample to air, then introduced the sample into the MBE system and annealed it for 5 min at 600 °C in UHV under arsenic pressure. Fig. 1 c) shows two AFM measurement of a-C covered GaAs surfaces for two different carbon deposition times. For too short carbon deposition (5 s) the GaAs surface is not fully covered by the a-C layer. When exposed to air these uncovered GaAs surface areas form a surface oxide layer. Annealing at approximately 600 °C in our MBE system leads then to typical deoxidation pits, which we also observe on bare GaAs surfaces after annealing (see Supplementary Fig. 2 b)). Growth on partially covered substrates is discussed in the Supplementary Note 2. In contrast, a completely a-C covered GaAs surface protects the GaAs surface from oxidation. This allowed us besides XRR measurements to precisely fine-tune the a-C deposition time to reach monolayer like thick a-C layers. We investigated the structure of the deposited carbon layers further and performed Raman spectroscopy measurements and X-ray photoelectron spectroscopy (XPS) measurement after sample annealing at 300 °C. Fig. 1 d) shows XPS-spectra in a binding energy range around the C1s carbon peak of two a-C layers deposited for 10 s and 300 s corresponding to an a-C thickness close to a monolayer and roughly (13±1.5) nm respectively together with a highly oriented pyrolytic graphite (HOPG) reference sample. The asymmetric C1s peak of the HOPG reference sample corresponds to $sp^2$-hybridized carbon. We observe no traces of impurities like oxygen or hydrogen for this HOPG reference sample. The measured spectrum of the monolayer like a-C layer ($t_{a-C}$=10 s) exhibits a similar spectrum and the same asymmetric peak shape thus revealing that our a-C deposition process indeed allows the fabrication of monolayer like thin $sp^2$-hybridized a-C layers. In contrast, the C1s peak shape of the (13±1.5) nm thick a-C layer ($t_{a-C}$=300 s) exhibits a slightly different peak shape due to the coexistence of $sp^2$- and $sp^3$-hybridized carbon in the layer. Note that the $sp^2$-hybridized contribution is still dominant. A detailed decomposition of the C1s peaks and the atomic ratios of observed elements for all samples is presented in Supplementary Note 3. Further XPS measurements confirm that the a-C layer protects the underlying GaAs surface from oxidation. There are only minimal $As_2O_3$ and $Ga_2O_3$ signals visible in the XPS spectrum obtained for H-Ar plasma cleaned substrates compared to reference measurements performed on GaAs substrates, which have not been plasma-treated (see Supplementary Fig. 3). The structure of the a-C layers was further investigated by Raman spectroscopy measurements exemplarily shown in Fig. 1 e). Two peaks at (1360±4) $cm^{-1}$ and (1587±2) $cm^{-1}$ are derived by two fitted Voigt-profiles. Please note that we do not observe the typical graphene 2D peak at roughly 2690 $cm^{-1}$ in our samples. Highly crystalline graphite exhibits a Raman-peak at 1580 $cm^{-1}$, diamond-like carbon only at 1330 $cm^{-1}$ and microcrystalline and disordered graphite two peaks at roughly 1580 $cm^{-1}$ and 1350 $cm^{-1}$ indicating that our carbon layers are mostly in a highly disordered graphitic or amorphous phase [25-28]. In summary, XPS and Raman measurements confirm that our deposited carbon layers are mostly graphitic/amorphous carbon with dominant $sp^2$-binding. This was reported as an important factor for successful remote epitaxy: In-plane $sp^2$-hybridized bonds only screen the substrate potential but do not modulate it in growth direction [29]. In summary, we want to highlight two aspects of our a-C fabrication process: First, $sp^2$-hybridized a-C layers are fabricated at room-temperature in a scalable, cheap and vacuum compatible process. Second, although we did not grow graphene, the a-C covered substrates are well suited as templates for high-quality remote epitaxy as discussed later for several examples. We have obtained very similar results for a-C layers fabricated on Si(001) and 3C-SiC(001) and think that the a-C layer fabrication process presented here can be adopted to any substrate that has decent surface roughness and can withstand temperatures of at least 300 °C tested here.

**Nucleation of GaAs on a-C covered GaAs**

After precisely tailoring the a-C layer thickness, we investigated as a model system for homoepitaxial growth by remote epitaxy the nucleation behavior of GaAs on a-C covered GaAs substrates. We performed GaAs nucleation on a-C covered GaAs substrates with a-C layer thicknesses close to the monolayer limit ($t_{a-C}$=10 s-45 s) at 300 °C and a III-V ratio of $\approx$ 30 (see Methods). We chose a low nucleation temperature due to the reduced substrate-layer interaction and the lower substrate wettability induced by the a-C layer [30, 31]. First, we discuss the surface morphology of GaAs layers grown on different a-C layers. Fig. 2 shows 1x1 µm$^2$ and 5x5 µm$^2$ AFM images of the surface morphology of GaAs layers with a nominal thickness of 2 nm grown on a-C layers deposited for a) 10 s, equivalent to a monolayer a-C, and b) 45 s. The images reveal the growth of individual islands, but a clear difference is that the a-C surface is nearly completely covered by the epitaxial GaAs layer for $t_{a-C}$=10 s, while there are uncovered areas for $t_{a-C}$=45 s a-C deposition. Obviously, the a-C layer alters the GaAs nucleation which is in clear contrast to GaAs growth on bare GaAs substrates for which we observe layer-by-layer growth and smooth surfaces (see Supplementary Fig. 2e)). We achieved growth of closed epitaxial layers at 300 °C for thicknesses >2 nm, which demonstrates that our low-temperature deposition allows to create a closed film on the amorphous carbon surface quickly. We think that the substrate potential damping induced by the a-C layer leads to a reduced surface wetting by the ad-atoms forcing Vollmer-Weber growth till the a-C layers are fully covered by GaAs [26]. Fig. 2 c)-e) shows AFM images of 25 nm thick GaAs layer grown at 300 °C on a-C layers deposited for c) 10 s, d) 20 s, e) 45 s. A general statement is, the thinner the a-C layer, the smoother the surface. The surface of the GaAs layer deposited on the monolayer thick a-C layer (10 sec) exhibits an RMS roughness of 0.4 nm, comparable to the GaAs surface RMS roughness of 0.3 nm measured for growth on bare GaAs substrates. Doubling the carbon deposition time (20 sec) results in a GaAs layer surface roughness of 0.68 nm and increasing the carbon deposition time again (45 sec) causes an even rougher surface morphology with an RMS roughness of 3.1 nm. Please note that the RMS roughness of the as-deposited a-C is for all layers ≤0.3 nm. The smooth GaAs surface morphology on 10 s a-C layers verifies the rapid coalescence of the initial nuclei at low growth temperature and for monolayer like thick a-C. A 10x10µm$^2$ AFM measurement of this sample (see Fig. 2 c)) verifies a smooth and flat surface with a surface roughness of 0.32 nm also on a large scale. This demonstrates that growth on our monolayer-like thin a-C covered substrates is suitable for the fabrications of optically and electrically active heterostructures requiring smooth interfaces in general. We attribute the increase of surface roughness with increasing a-C deposition time to a stronger substrate potential screening of thicker a-C layer. Interestingly, we do not observe a clear boundary for the remote epitaxial growth of GaAs on a-C in contrast to reported growth on transfer or epitaxial graphene where GaAs remote epitaxial growth was observed on 1 ML graphene but polycrystalline growth on ≥2 ML graphene [6,11]. We speculate that during the a-C deposition the GaAs is exposed to C- and H-radicals and ion bombardment which forces interactions of C- and H-atoms with the GaAs surface resulting probably in the formation of an interlayer or an a-C layer directly on the GaAs surface forcing direct charge transfer between GaAs substrate and a-C layer [29]. In contrast, there is only minor interaction of transferred graphene with a host substrate. Further, a-C layers are not deposited in a layer-by-layer fashion, but randomly forming an amorphous phase which is in contrast to monolayer-by-monolayer stacking of graphene. Finally, the PECVD of a-C layers allows to grow arbitrary thick a-C layers and perform remote epitaxy in a certain range up to roughly 2 nm for which the substrate-layer interaction is successively reduces but still transferring the substrate crystal orientation to the growing film. However, successively increasing the a-C thickness results in surface degradation and as discussed below also favors the formation of crystal defects. TEM (transmission electron microscopy) measurements were performed to gain deeper insights in the atomic structure. Fig. 2 f) shows an Annular Dark Field (ADF) STEM image of the GaAs/a-C/GaAs interface with an a-C layer deposited for 10 s. The GaAs substrate crystal structure is transferred to the GaAs epitaxial layer proving high quality crystal growth. Further, a darker contrast was observed at the interface. A distinct interaction gap as reported in [23] between GaAs layer and substrate is not observed here, neither carbon was detected by EDX scans along the interface (not shown here), which contrasts with our XRR, XPS and Raman spectroscopy measurements presented above. However, Geometric Phase analysis applied to ADF images revealed strain at the interface along the growth direction indicated by the red color at the interface. We speculate that, although we did not directly detect carbon by TEM measurements at the interface, this lattice stretching originates from the present of our deposited a-C layer. Similar to remote epitaxy on graphene, the substrate-layer binding is altered causing a different nucleation behavior on the a-C-substrate surface. While remote epitaxy on graphene induces a clear interaction gap, deposition of an a-C layers causes a tunable lattice stretching. We think that this lattice stretching appears mainly for an a-C thickness where the leaking potential of the substrate guides the nucleation on the amorphous carbon layer, i.e., for monolayer-like thin amorphous carbon thickness. In this case, the separation of substrate and layer is small enough to translate the crystal structure/potential through the amorphous carbon, similar to reported remote epitaxy on graphene [6,8]. For thicker amorphous carbon layers, the leaking potential is screened by the amorphous carbon layer thus impeding translation of the substrate crystal lattice to the film resulting in defective crystal film growth on the amorphous carbon.

We investigated the crystal structure of epitaxial layers on large scale with different aforementioned a-C thicknesses in more detail by HR-XRD measurements. Most interesting are epitaxial layers grown on monolayer like thin a-C layers ($t_{a-C}$=10 s) as for these layers the substrate-layer interaction is assumed to be strongest thus the best crystal quality is expected. To distinguish between the reciprocal lattice reflection of substrate and layer, we grow 25 nm thick $In_xGa_{1-x}As$ (x<0.1) layers on a-C covered deposited for 10 s, 20 s, 45 s on GaAs substrates. The epitaxial $In_xGa_{1-x}As$ (x<0.1) layers are pseudomorphic for such low indium concentrations which we confirmed by reciprocal space maps (RSMs) of the ($\bar{2}\bar{2}4$)-reciprocal lattice reflection (not shown here). Fig. 3 a) shows $\omega$-$2\theta$- and off axis $\phi$-scans for a wide-angle range. A sharp (002)- and (004)-reciprocal lattice reflection of substrate and $In_{0.09}Ga_{0.91}As$ layer grown on monolayer like a-C ($t_{a-C}$=10 s) in the $\omega$-$2\theta$-scans reveals transfer of the substrate crystal orientation to the layer. For $In_{0.06}Ga_{0.94}As$ layers grown on a-C layers deposited for $t_{a-C}$=20 s we observe a weaker layer peak and for $t_{a-C}$=45 s there is no layer peak observed, which points to reduced crystal quality of the epitaxial layer with increasing a-C layer thickness. The $\phi$-scans reveal a clear 90° peak periodicity. The absence of any other reflections in both, the $\omega$-$2\theta$-and $\phi$-scans, indicate the absence of polycrystalline grains. To gain more information about the crystal quality of the $In_xGa_{1-x}As$ (x<0.1) layers, RSMs of the (004)-reciprocal lattice reflection were performed. Fig. 3 b) shows exemplarily the (004)-RSM for a 25 nm thick $In_{0.09}Ga_{0.91}As$ layer grown on a monolayer like a-C layer ($t_{a-C}$=10 s). Pendellösung fringes are observed indicating a smooth surface. We performed an $\omega$-scan of the (004)-reciprocal lattice reflection of the layer. The FWHM $\Delta\omega = (0.013\pm0.001)$ ° corresponds to an estimated dislocation density of $(3.58 \pm 0.06) \times 10^6$ cm$^{-2}$ (see Supplementary Note 4) [32]. The GaAs substrate exhibit a FWHM $\Delta\omega = (0.006\pm0.001)$ ° in the used set-up. These results demonstrate high crystal quality remote epitaxial growth on a-C layers, i.e. the dislocation density is similar to values we expect for growth on bare GaAs. Probably our monolayer-like a-C thickness and the ultra-smooth surface of our a-C layer are crucial to prevent the formation of dislocations originating from the a-C layer. Increasing the a-C deposition time, thus decreasing the substrate layer interaction, forces the degradation of the epitaxial layer. We investigated the crystal quality of the 25 nm pseudomorphic $In_{0.06}Ga_{0.94}As$ layer grown on the thicker a-C layer ($t_{a-C}$=20 s) (see Fig. 3 c)) and observe that the FWHM of the (004)-reciprocal lattice reflection increases to $\Delta\omega = (0.72\pm0.02)$ ° thus the layer quality degrades strongly. The corresponding estimated dislocation density is $(1.1 \pm 0.06) \times 10^{10}$ cm$^{-2}$ (see Supplementary Note 4) [32]. The dislocation density is very sensitive to the a-C layer thickness. We performed comparative TEM measurements (see Supplementary Note 4 and Supplementary Fig. 4) of a 25 nm thick $In_{0.09}Ga_{0.91}As$ layer grown on a-C covered GaAs ($t_{a-C}$=10 s). We estimate a stacking fault density of the investigated area of >$10^{10}$ cm$^{-2}$. This density is larger than the dislocation density of <$10^7$ cm$^{-2}$ obtained from our HR-XRD measurements for a similar layer structure. We speculate that the a-C layer thickness might vary locally on nanometer scale thus causing a locally varying dislocation density. As discussed above the dislocation density is very sensitive to the a-C layer thickness so the strong increase of the dislocation density might be related to a slightly thicker a-C layer for the investigated TEM cross section area. Further, the TEM measurements probe a small cross section area (30x300 nm$^2$), while the HR-XRD measurements average over a much larger area (roughly 10x10 cm$^2$). Our results demonstrate that remote epitaxy on a-C is possible up to a thickness of roughly 2 nm of the a-C layer, but smooth and high crystal quality layers are obtained for monolayer-like thin a-C layers. The degradation of the layer crystal quality exhibits the same trend as the roughening of the surface morphology with increasing a-C layer thickness discussed above.

Next, we analyze the influence of different ad-atom species and the substrate growth temperature on the epitaxy on a-C covered GaAs surface. The initial state of growth is governed by the substrate and layer ionicity, the kinetic processes of the ad-atoms on the a-C covered surface and the wettability of the a-C covered surface [7,8,30]. We compare here the nucleation of GaAs and AlAs on a-C covered GaAs substrates for 300 °C to get insight in the influence of adsorption and migration energies of different materials [7] while keeping the substrate-layer lattice mismatch negligible. Fig. 3 d) shows an 1x1 µm$^2$ AFM image of the surface of a 2 nm thick AlAs layer grown on a-C covered GaAs at 300 °C. A smooth surface with a roughness of 0.3 nm is obtained for this 2 nm thin AlAs layer, which is smoother compared to the grainy surface obtained for growth of 2 nm thick GaAs nucleation layers already presented in Fig. 2 a). The energy barrier for Al atom migration on the a-C covered GaAs is probably larger compared to that of Ga so that AlAs covers the a-C layer better compared to GaAs because both tend to not wet the surface but forming small islands on the surface. We conclude, that the adatom species plays an important role in the nucleation on a-C. For rapid and smooth surface coverage, low-temperature AlAs nucleation should be performed on a-C covered GaAs substrates. We further investigated the growth on a-C at elevated temperatures. Fig. 3 e) shows an 1x1 µm$^2$ AFM image of a 25 nm thick GaAs layer grown on a-C covered GaAs at 500 °C. In comparison to the low-temperature growth at 300 °C presented in Fig. 2 c), the layer exhibits a grainy surface, which seems to consist of successively merged islands The growth temperature has a major impact on the ad-atom migration and re-evaporation thus governing the initial stage of growth on a-C covered GaAs substrates. This contrasts with growth on bare GaAs substrates where we observe smooth growth of GaAs or AlAs films in a wide temperature range of roughly 300 °C-650 °C with our growth parameters. Deposition of individual GaAs islands on a-C covered GaAs was also observed in [30] for high temperature nucleation at 650 °C. The strong change of the surface morphology with increasing growth temperature seems reasonable as the competing processes on the surface (ad-atom migration, re-evaporation, nucleation) very probably follow an exponential temperature dependence of thermally activated processes. A high substrate temperature enhances the ad-atom re-evaporation from a-C surfaces and forces the successive growth of existing nuclei on the a-C covered GaAs surface [9]. To achieve a smooth layer with homogeneous a-C coverage, the growth temperature should be kept low during the initial stage of growth on a-C layers. For applications, Al-rich nucleation layers are probably preferred as they form closed and smooth surfaces for lower film thicknesses.

The fabrication of heterostructures and devices usually requires thicker films than discussed so far, so we performed subsequent overgrowth of our thin nucleation layers and obtained smooth Al- and Ga-rich films (see Supplementary Fig. 5 a)-b)). Following the epitaxial lift-off approach presented in [7], we performed the epitaxial lift-off of thicker epitaxial layers (see methods). A photograph and optical microscopy image with large magnification (see Supplementary Fig. 5 c)) of a 300 nm thick $In_{0.1}Ga_{0.9}As$ film lifted-off from the substrate reveals that the entire transferred film is free of cracks; beside the lower edge where we touched the film during lift-off with a tweezer. AFM, $\theta/2\theta$- and Raman measurements clearly reveal the successful lift-off of the $In_{0.1}Ga_{0.9}As$ film (see Supplementary Note 5). Fig. 4 shows XRR, Raman and AFM measurements of the substrate and the $In_{0.1}Ga_{0.9}As$ film after lift-off, respectively. While Raman measurements on the backside of the film show no traces of a-C, we observe a clear carbon-related signal from the substrate after layer lift-off. This observation is strengthened by the XRR measurements performed on the substrate after lift-off. Compared to the XRR spectra of a bare GaAs substrate, the XRR spectra of the GaAs substrate after layer lift-off shows a clear oscillation. Both, the Raman and XRR measurement of the substrate after lift-off are nearly identical to the spectrum obtained after freshly preparing an a-C layer on the GaAs substrates (see Fig. 1 a) and d)) revealing that the a-C layer stays on the substrate after layer lift-off. We investigated the surface morphology of the a-C layer on the substrate after layer lift-off. The 5x5μm² AFM image in Fig. 4 c) reveals a smooth a-C surface exhibiting an RMS roughness of 0.3 nm similar to a freshly prepared a-C covered GaAs substrate (as presented in Fig. 1 c)). In conclusion, our measurements show that the a-C layer remains on the substrate after lift-off thus potentially allowing wafer recycling and reduction of production cost and also pave the way towards assembling different membranes allowing versatile device fabrication [3,7].

**Growth on a-C covered InP substrate**

In a next step, we transferred our approach from our model system GaAs to material systems, which require more sensitive growth parameters. First, we discuss the growth of pseudomorphic $In_xGa_{1-x}As$ layers on InP substrates demonstrating the flexibility of our a-C preparation method. In previous works, remote epitaxy has been reported for a-C and BN layers, respectively, fabricated at elevated temperature of ≥680 °C on $Al_{0.5}Ga_{0.5}As$ and SiC substrates [3,30]. Such high temperatures are obviously not compatible with temperature sensitive substrates like InP exhibiting a lower decomposition temperature of roughly 550 °C. In contrast, with our room-temperature PECVD method, we can prepare a-C layers on temperature sensitive substrates as InP. We covered InP substrates with a-C deposited for 10 s and performed subsequent epitaxial overgrowth of nearly lattice matched 25 nm thick $In_{0.5}Ga_{0.5}As$ layer at 300 °C with a III-V ratio of ≈ 30. The inset of Fig. 5 a) shows exemplary an 1x1 μm² AFM image of the layer surface revealing a smooth layer similar to layers grown on bare InP substrates. To gain deeper insight in the crystal structure, HR-XRD measurement were performed. The $\omega$-$2\theta$- and $\phi$-scan in Fig. 5 a) and b) reveal the absence of polycrystalline grains in the layer and the (001)-crystal orientation of layer and substrate. Fig. 5 c) and d) shows the RSMs of the (004)- and ($\bar{2}\bar{2}4$)-reciprocal lattice reflection of a more mismatched $In_xGa_{1-x}As$ layer respectively revealing a pseudomorphic layer for which we derive the indium concentration to x=(65±1) %. Pendellösung fringes in both RSMs indicate a smooth surface of the $In_{0.65}Ga_{0.45}As$ layer, similar to the $In_{0.09}Ga_{0.91}As$ nucleation layer grown on an a-C covered GaAs substrate presented above. We performed an $\omega$-scan of the (004)-reciprocal lattice reflection of the $In_{0.65}Ga_{0.45}As$ layer. The FWHM $\Delta\omega = (0.015\pm0.001)$ ° corresponds to an estimated dislocation density of $(4.76 \pm 0.06) \times 10^6$ cm$^{-2}$ (see Supplementary Note 4) [32]. The InP substrate exhibit a FWHM $\Delta\omega = (0.005\pm0.001)$ ° in the used set-up. Further, we fabricated droplet etched In$_{0.57}$Ga$_{0.43}$As quantum dots embedded in an In$_{0.52}$Al$_{0.48}$As matrix grown on an a-C covered InP substrate (see Supplementary Note 6) and performed photoluminescence measurements at 16 K and 4.5mW excitation power at 635nm (see Supplementary Fig. 6). A broad peak in the range of 1350-1600 nm is clearly visible in the spectrum revealing successful growth of an optically active structure on our a-C covered InP substrates. We attribute the broad peak shape to the non-optimised growth parameters on the a-C covered InP substrate, i.e. the quantum dot size distribution might be large resulting in broad emission. To our knowledge these are the first demonstrations of remote epitaxy of high crystal quality layers grown on a-C covered InP substrates. Consequently, using an a-C interlayer on InP substrates paves a way for wafer-scale lift-off from a-C covered InP substrates and the fabrication of freestanding In$_x$Ga$_{1-x}$As layers from InP substrates. This might be of great interest for the fabrication of freestanding heterostructures operating in the optical-C band because lifting-off membranes from InP substrates is quite challenging due to the lack of an established epitaxial lift-off process.

**Growth of metastable cubic AlN and GaN**

Remote epitaxy of group-III nitrides has been investigated intensively and it was shown that the thermal stability and large substrate ionicity compared to the group-III-arsenides is favorable for remote epitaxial growth. It was mostly reported on the remote epitaxy of group-III nitrides in the wurtzite crystal structure [8,12-14, 21,22] but Littmann et al. [33] discussed remote epitaxy of thick GaN layers grown in the metastable cubic crystal structure, which is especially challenging. Growth of cubic nitrides in the metastable phase is challenging due to a narrow growth parameter window and absence of lattice matched substrates and serves as a sensitive test candidate for growth on our a-C layers. The growing crystal is forced into the cubic (c-) structure by the substrate so reducing the substrate-layer interaction impedes the growth of cubic nitrides. Littmann et al. discussed remote epitaxy of c-GaN on transfer-graphene and pointed out that remote epitaxy suffers from the growth of hexagonal GaN inclusions at graphene grain boundaries, wrinkles and pinholes [33]. Further, the growth temperature range is delicate for cubic nitride growth: c-AlN and c-GaN are grown at $\geq$760 °C and the substrate temperature cannot be lowered for nucleation as employed for GaAs growth. We first discuss the nucleation of thin c-AlN layers grown on a-C covered c-AlN buffer layers grown on 3C-SiC/Si(001) substrates at 760 °C. First, a few monolayer thin c-AlN layer was grown on 3C-SiC/Si(001) substrates to exploit the large ionicity of 0.449 [36] of the c-AlN layer for the remote epitaxial overgrowth. Then, this stack was covered by a-C deposited with monolayer-like thickness following subsequent overgrowth by a c-AlN layer at 760 °C. Note that the substrate-layer mismatch between 3C-SiC(001) and c-AlN is 0.3 % so we can assume a pseudomorphic c-AlN layer on both, the 3C-SiC/Si(001) substrate and the a-C covered 3C-SiC/Si/AlN quasi-substrate. Fig. 5 g) shows exemplarily a 1x1 µm$^2$ and a 10x10 µm$^2$ AFM image of an approximately 5 nm thick c-AlN layer grown on an a-C layer (well below the critical thickness). We want to highlight two points here. First, the c-AlN layer exhibits a closed surface with an RMS roughness of 0.45 nm on 1x1 µm$^2$ revealing proper nucleation of c-AlN on the a-C layer. Second, we do not observe any cracks or surface defects in the film. This demonstrates the remote epitaxy of metastable crystals on a-C and also the temperature stability of a-C. Following these promising results we performed more delicate mismatched growth (~3 %) of roughly 320 nm thick c-GaN layers at 845 °C on the a-C covered 3C-SiC/Si(001) substrate. Fig. 5 e) shows a RHEED image measured in [110] direction of the c-GaN layer grown on a-C together with the RHEED pattern of the bare a-C covered substrate for comparison. The amorphous structure of the deposited a-C film leads to a diffuse pattern as the streaky pattern of the substrate interferes with the diffuse signal from the a-C layer. Subsequent overgrowth forces intense streaks typical for growth of c-GaN on 3C-SiC/Si(001) substrates: The slightly spotty reflections originate from electron scattering on islands while the long streaks indicate formation of a 2D surface reconstruction. Besides that, hexagonal reflections are not observed indicating dominant c-GaN growth [35]. To quantify this in more detail, we performed HR-XRD measurements. Fig. 5 f) shows the RSM of the (002)-c-GaN reciprocal lattice reflection. An intense reflection originating from the (002)-c-GaN reflections and a much weaker reflection of hexagonal inclusions can be identified. The hexagonal inclusion fraction of roughly 16 % is calculated from the integrated intensities of the cubic and hexagonal reflection [35]. This value is significantly lower than the 23% h-GaN reported by Littmann et al. for optimized growth parameters and we attribute this to the superior quality of our a-C layer in contrast to transferred graphene used by Littmann et al. [33]. Their layers suffered from large cracks hosting h-GaN grains which they attributed to wrinkles or folds of the transfer graphene. We think the reduction of hexagonal inclusion is a strong indicator of the large-scale integrity of our a-C layer. Optimization of the delicate growth parameter of the c-GaN growth, especially the nucleation on a-C probably could reduce the number of hexagonal inclusions further, but this is beyond the scope of this paper. To our knowledge this is the first demonstration of wafer scale remote epitaxial growth of crack or large area defect free cubic group-III-Nitrides. Our results pave the way to lift-off cubic group-III nitride layers and allows the fabrication of free-standing cubic group-III nitride films which is nowadays quite challenging due to chemical robustness of the group-III-nitrides and the 3C-SiC(001) pseudosubstrate.

**Growth of highly mismatched InGaAs layers on a-C covered GaAs**

In this sub-section, we discuss the initial stage of remote epitaxial growth of $In_xGa_{1-x}As$ layers ($0.48 \leq x \leq 0.52$) on a-C covered GaAs as a model system for large substrate-layer mismatch (roughly 3.6 %). This material system is also of technological interest because such indium concentration allows the fabrication of heterostructures emitting in the technological important optical C-band and integration into the GaAs material system allowing the in-situ growth of distributed Bragg reflectors. Fig. 6 a) shows two AFM images of 25 nm thick $In_{0.50}Ga_{0.50}As$ layers grown at 300 °C and a III-V ratio of $\approx 30$ on a-C deposited for 10 s and for 20 s, respectively. Both surfaces exhibit a grainy morphology and an RMS surface roughness of 0.7 nm and 1.7 nm, respectively. Increasing the a-C layer thickness results in rougher surfaces like for GaAs growth on a-C covered GaAs discussed above. However, the surface is significantly rougher compared to the GaAs layers grown on a-C covered GaAs or the lattice matched $In_{0.50}Ga_{0.50}As$ layers grown on a-C covered InP. We conclude that the substrate-layer lattice mismatch and the layer relaxation play a crucial role during material nucleation on a-C layers. We investigated the crystal structure in more detail by performing HR-XRD measurements (see Fig. 6 b)-d)). The $In_xGa_{1-x}As$ layer is free of polycrystalline grains and substrate and layer are oriented in the (001)-crystal direction proven by $\omega$-$2\theta$- and $\phi$-scans presented in Fig. 6 c) and d). Further, RSMs of the (004)- and ($\bar{2}\bar{2}4$)- reciprocal lattice reflection of $In_xGa_{1-x}As$ layers ($0.48 \leq x \leq 0.52$) grown on a-C covered GaAs deposited for 10 s and 20 s are shown in Fig. 6 b) together with $In_xGa_{1-x}As$ layers ($0.48 \leq x \leq 0.52$) grown with the same growth parameters on bare GaAs substrates. We derived the degree of relaxation from the ($\bar{2}\bar{2}4$)-RSMs and the FWHM of the $In_xGa_{1-x}As$ layer reflection from $\omega$-scans. All values are summarized in Tab. 1 together with the layer's dislocation densities which were calculated using the FWHM following [32] as shown already above. In comparison to growth on bare GaAs, growth on a-C leads to a stronger layer relaxation of +18 % while the dislocation density is similar. Further, the thicker a-C layer (20 s) causes stronger layer relaxation but also a larger dislocation density which is probably attributed to a stronger substrate screening. We performed growth of a 200 nm thick $In_xGa_{1-x}As$ layer on a-C covered GaAs as well as on bare GaAs and observed a similar trend. While the $In_{0.52}Ga_{0.48}As$ layer grown on a-C is nearly fully relaxed, the $In_{0.50}Ga_{0.50}As$ layer grown on bare GaAs exhibit -11 % relaxation. Both layers exhibit a similar dislocation density. Our results show that $In_xGa_{1-x}As$ layers (0.48≤x≤0.52) grown on a-C are stronger relaxed compared to similar layers grown on bare GaAs while showing similar dislocation densities. Such a relaxation behavior was also observed for $In_xGa_{1-x}As$ growth on transfer graphene but larger dislocation densities were measured attributed to a low transfer graphene quality [39]. Considering the high crystal quality and smooth surface morphology of the $In_{0.5}Ga_{0.5}As$ grown on InP, it is likely that strain is a key factor for the comparatively/rather high defect density and grainy surface of the $In_xGa_{1-x}As$ (0.48≤x≤0.52) layer grown on a-C covered GaAs. As a layer relaxation mechanism besides the formation of dislocations on graphene covered substrates, a layer slip over the graphene was discussed [18,39,40]. Due to a weaker substrate-layer bonding, the layer can partially release strain across this interface. We think that this process could explain the 18% difference in the degree of relaxation between layers grown on a-C covered and bare GaAs. The overall large dislocation density of the grown layer is very probably due to the large lattice misfit of 3.6 % because we do not observe it for the lattice matched layers discussed above. For such large mismatch, the corresponding critical layer thickness is only a few nanometer so that relaxation sets in before the film covers the entire a-C film. As already shown, the $In_xGa_{1-x}As$ (0.48≤x≤0.52) layers growth starts in the Vollmer-Weber mode on the a-C layer and the a-C coated substrates is not covered completely before the critical thickness is reached. The individual islands are strained and maybe the strain is relaxed by both, the formation of dislocations in the initial islands and by a slip of the islands over the a-C layer. Finally, we would like to stress that our results show superior $In_xGa_{1-x}As$ (0.48≤x≤0.52) quality on a-C covered InP compared to growth on a-C covered GaAs.

**Conclusion and Outlook**

We demonstrated the direct low-temperature deposition of ultrathin 2D a-C layers on III-V semiconductors by PECVD as a universal template for remote epitaxy. A route to fabricate thin a-C layers on various sensitive III-V host substrates enabling subsequent remote epitaxial overgrowth was presented. X-ray photoelectron and Raman spectroscopy measurements reveal that the layers consist predominately of $sp^2$-hybridised amorphous carbon. Precisely tailoring the a-C layer thickness down to monolayer thickness via the deposition time allows superior tunability of the substrate-layer interaction. For remote epitaxy, a low-temperature nucleation step is favorable for nucleation of III-V material growth on substrates coated with thin a-C layers and that lattice matched growth leads to high crystal quality, low dislocation densities ($<1 \times 10^7$ cm$^{-2}$) and smooth surfaces with roughness <0.5 nm, demonstrating the flexibility of our carbon deposition process for temperature sensitive substrates like GaAs or InP and metastable cubic group-III nitride layer growth. Layer release from the substrate is demonstrated allowing the fabrication of versatile heterostructures from different membranes. Future investigation should include more detailed investigation of the atomic arrangement at the substrate-a-C-layer interface.

**ACKNOWLEDGMENT**

The authors like to acknowledge financial support by the Deutsche Forschungsgemeinschaft (DFG, German Research Foundation)– SFB-Geschäftszeichen TRR142/3-2022 – Projektnummer 231447078 Project B08 and C09

**Data availability**

All data supporting the findings of this study are available within the article and Supplementary Information files. They are also available from the corresponding authors upon reasonable request.

**Author contributions**

T.H. performed the remote epitaxy template fabrication, group-III arsenide growth, sample characterization, data interpretation and wrote the first draft of the manuscript. T.H. and D.R. conceived the core idea of the paper. P.M. and D.J.A performed growth and analyzation of cubic group-III nitride layers, respectively. A.W. performed XRR measurements and layer lift-off. D.D performed growth and photoluminescence measurements of droplet-etched quantum dots on a-C covered InP substrates. M.V. and G.G performed and interpreted XPS measurements and L. R. performed Raman measurements. A.M.S prepared, conducted and interpreted the TEM measurements. D.R. supervised the overall project. All authors critically revised the manuscript.

**Competing interests**

The authors declare no competing interests.

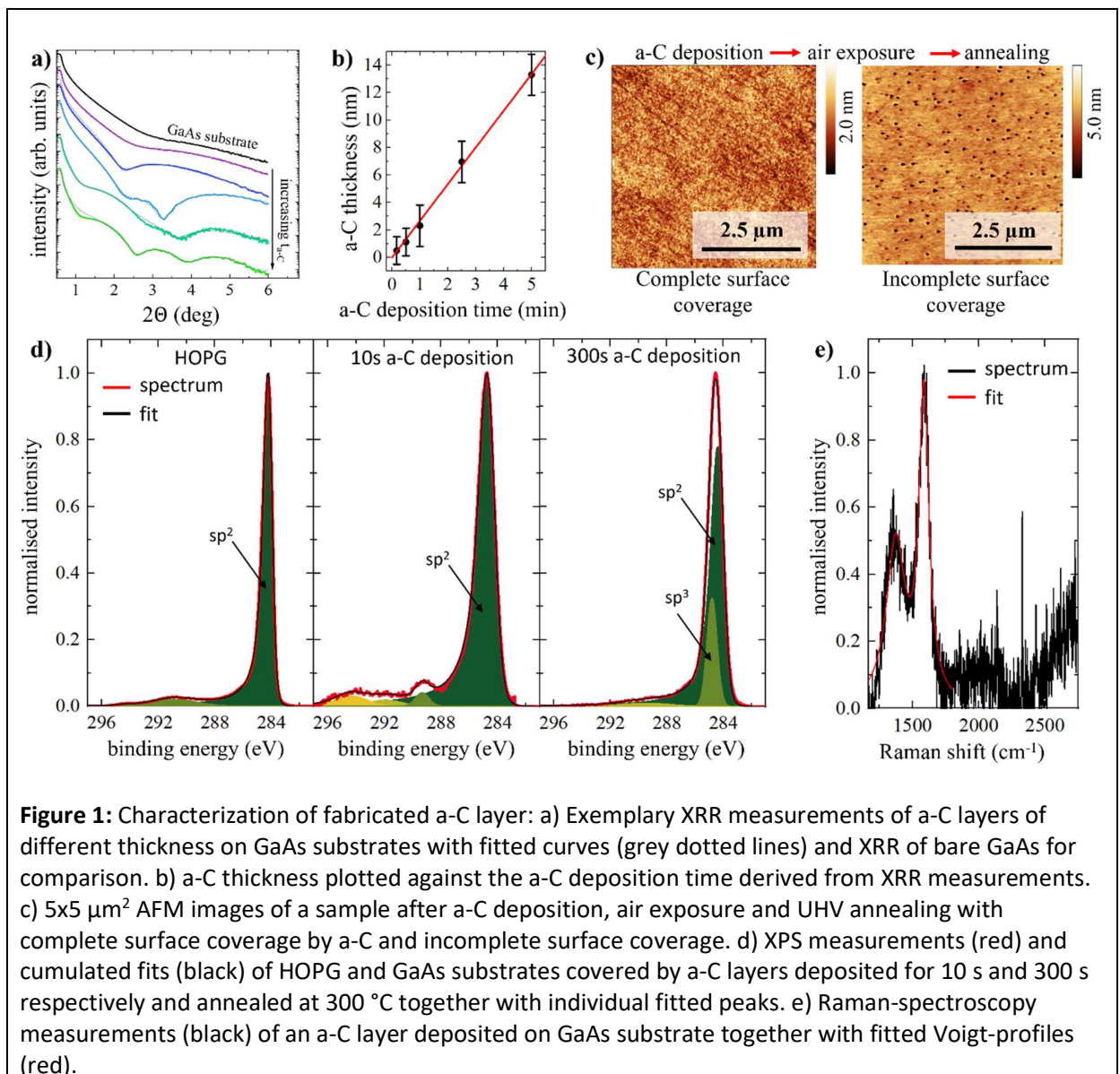

**Figure 1:** Characterization of fabricated a-C layer: a) Exemplary XRR measurements of a-C layers of different thickness on GaAs substrates with fitted curves (grey dotted lines) and XRR of bare GaAs for comparison. b) a-C thickness plotted against the a-C deposition time derived from XRR measurements. c) 5x5 µm² AFM images of a sample after a-C deposition, air exposure and UHV annealing with complete surface coverage by a-C and incomplete surface coverage. d) XPS measurements (red) and cumulated fits (black) of HOPG and GaAs substrates covered by a-C layers deposited for 10 s and 300 s respectively and annealed at 300 °C together with individual fitted peaks. e) Raman-spectroscopy measurements (black) of an a-C layer deposited on GaAs substrate together with fitted Voigt-profiles (red).

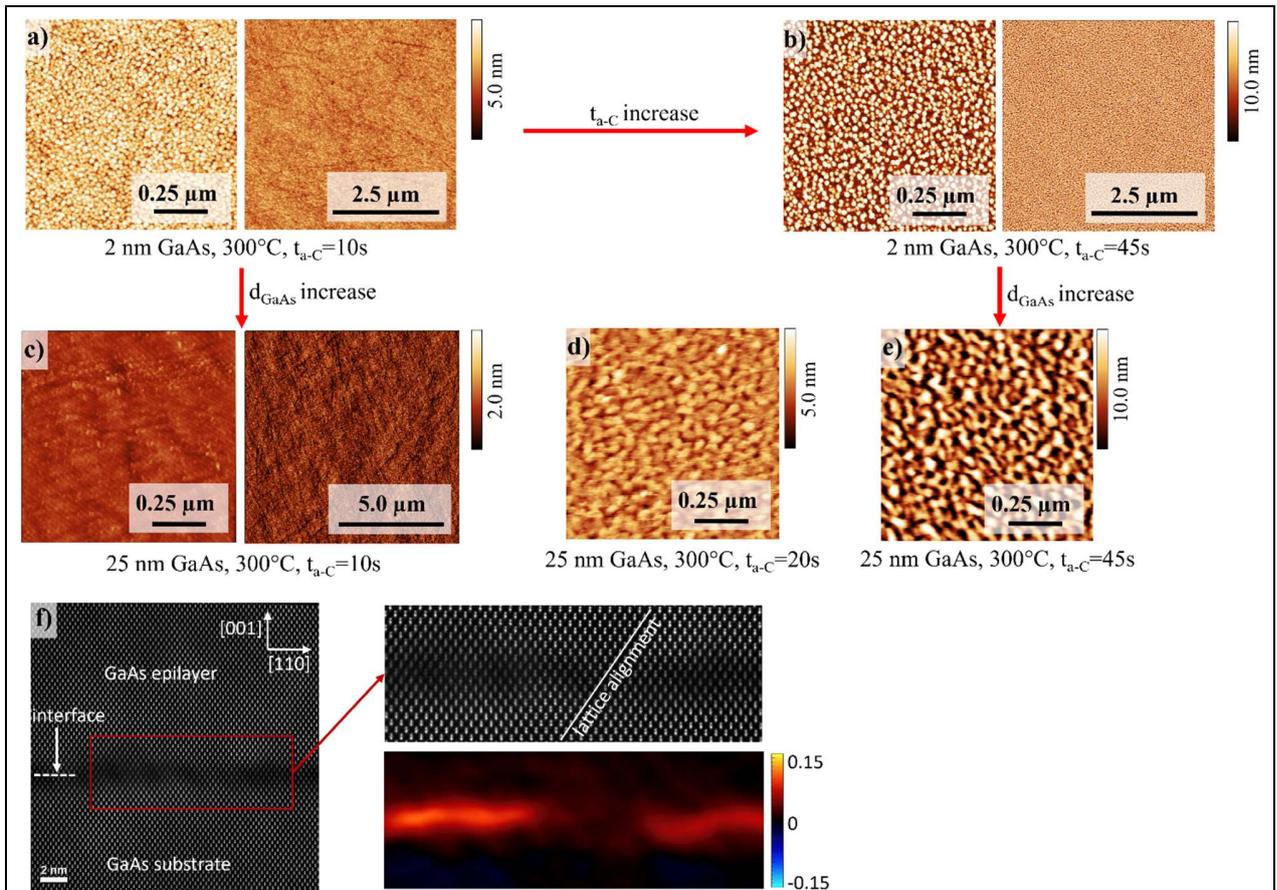

**Figure 2:** a)-e) 1x1 μm², 5x5 μm² and 10x10 μm² AFM image of 2 nm and 25 nm thick GaAs layers deposited at 300 °C on a-C covered GaAs. f) ADF-STEM TEM image of the GaAs substrate/layer interface together with a Geometric Phase Analysis.

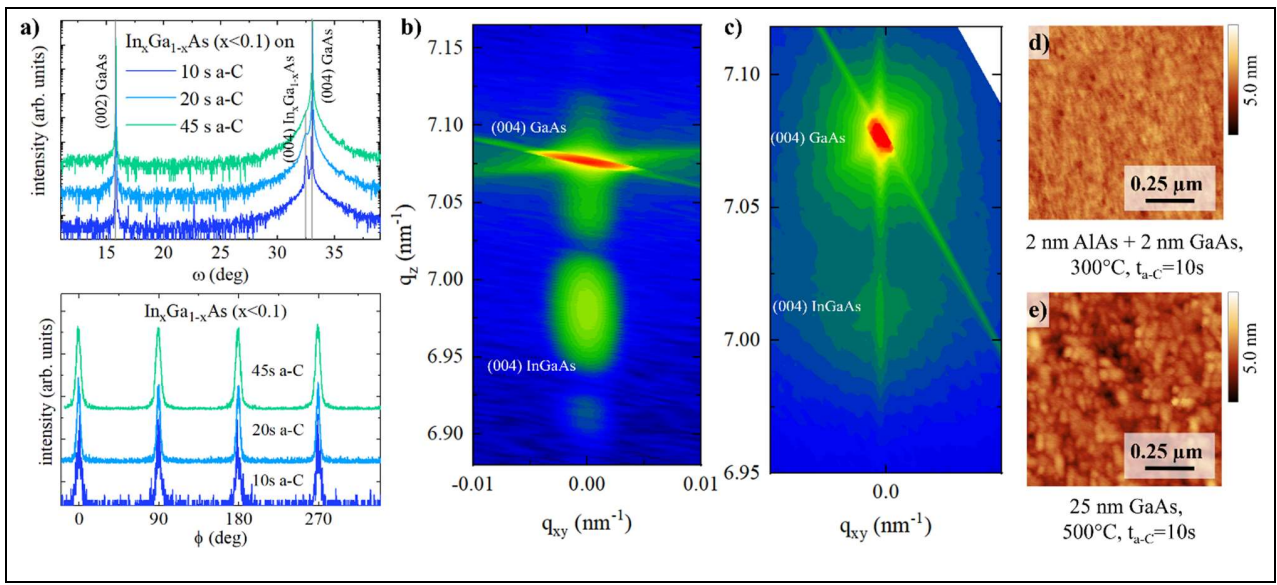

**Figure 3:** a) $\omega$-$2\theta$- and $\phi$-scan of 25 nm thick $In_xGa_{1-x}As$ (x<0.1) layer on a-C covered GaAs substrates ($t_{a-c}$=10 s, 20 s, 45 s). b)-c) Reciprocal space map of the (004)-reciprocal lattice reflection of a 25 nm thick $In_{0.09}Ga_{0.91}As$ and $In_{0.06}Ga_{0.94}As$ layer grown on a-C covered GaAs with an a-C layer deposition time of 10 s and 20 s respectively. d) 1x1 µm² AFM image of the surface of a 2 nm thin AlAs layers covered by 2 nm GaAs grown at 300 °C on a-C covered GaAs. e) 1x1 µm² AFM image of the surface of a 25 nm thick GaAs layer grown at 500 °C on a-C covered GaAs.

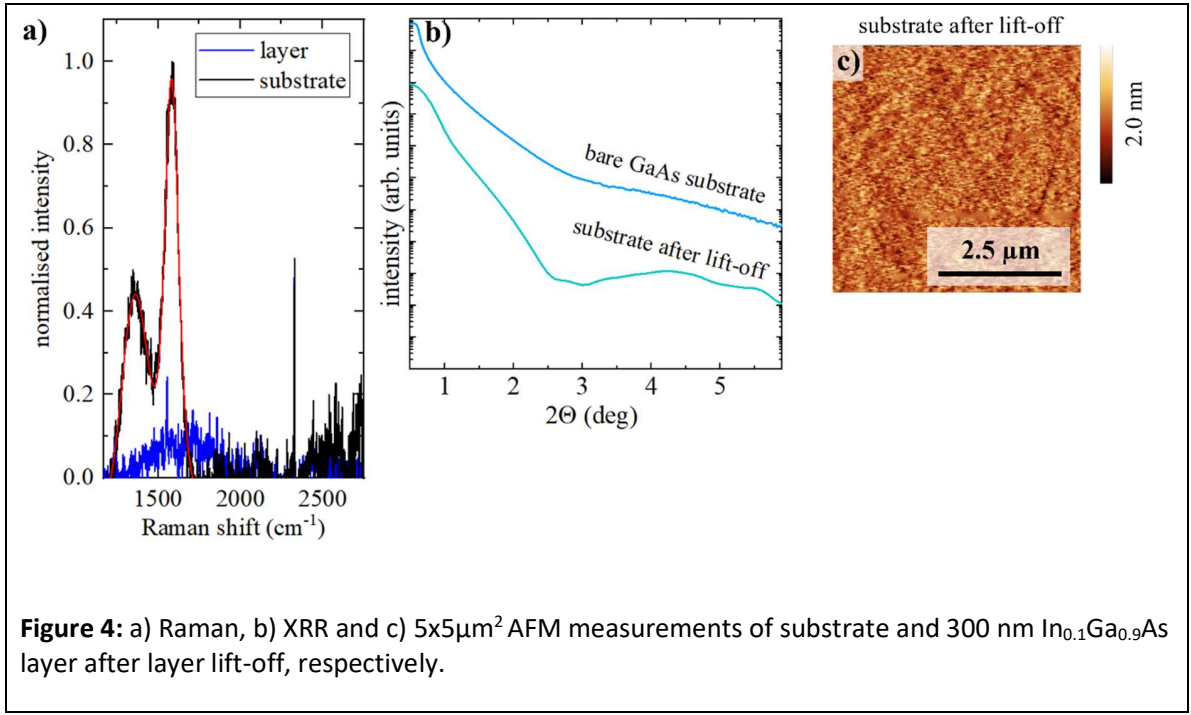

**Figure 4:** a) Raman, b) XRR and c) 5x5µm² AFM measurements of substrate and 300 nm $In_{0.1}Ga_{0.9}As$ layer after layer lift-off, respectively.

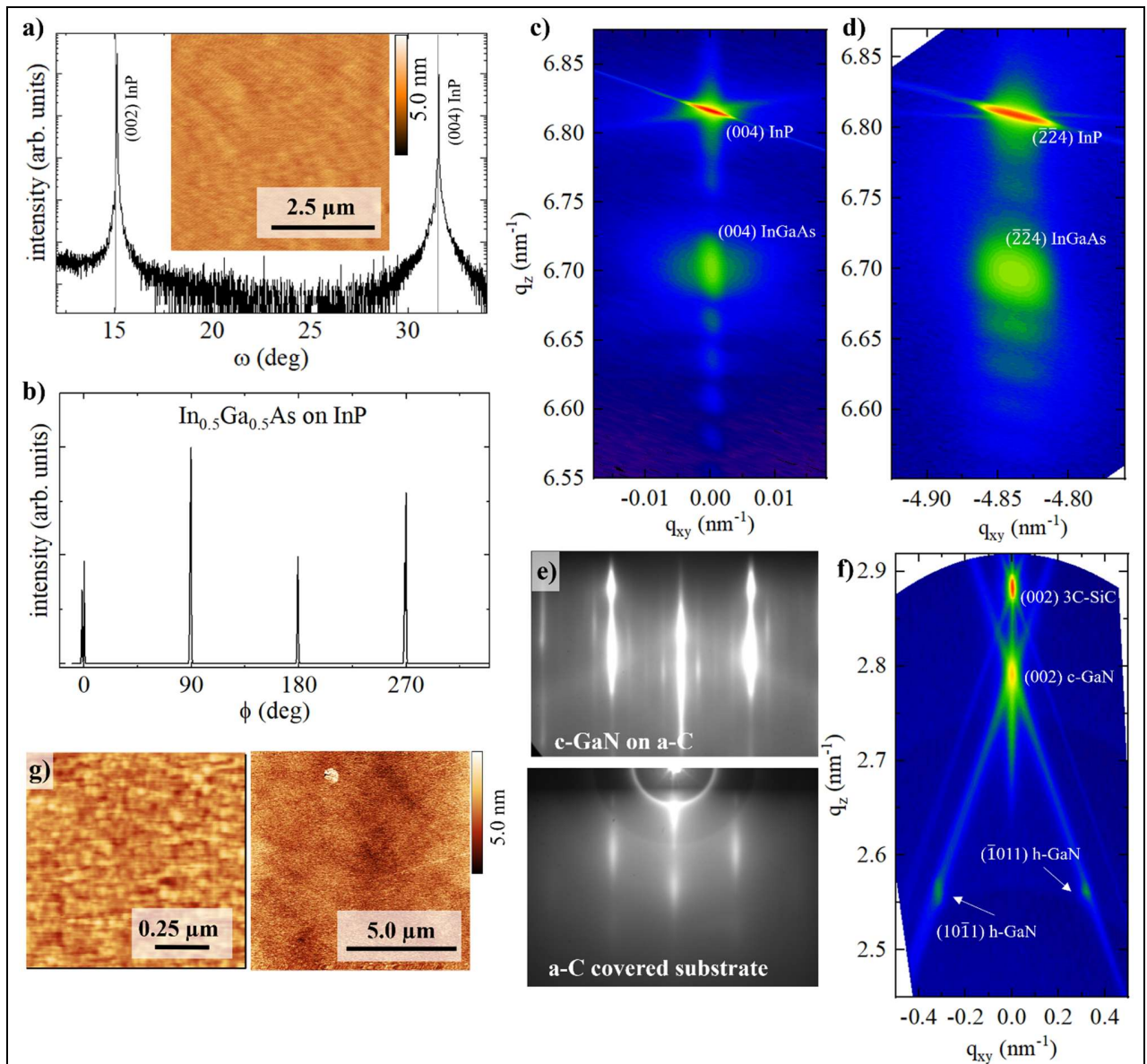

**Figure 5:** a)-c) In$_x$Ga$_{1-x}$As layer grown on InP substrate. d)-f) c-AlN and c-GaN layer grown on 3C-SiC/Si(001) quasi-substrate. a) $\omega$-2$\theta$-scan and b) $\phi$-scan of a 25 nm thick In$_{0.50}$Ga$_{0.50}$As layer and a 1x1 µm² AFM image of the surface (inset) grown on a-C covered InP at 300 °C. c)-d) reciprocal space map of the (004)- and ($\bar{2}\bar{2}4$)-reciprocal lattice reflection of a 25 nm thick In$_{0.65}$Ga$_{0.45}$As layer grown on a-C covered InP at 300 °C. e) RHEED images of a 320 nm thick c-GaN layer grown at 845 °C on a-C covered 3C-SiC/Si(001) quasi-substrate and on the bare quasi-substrate for comparison. f) RSM of the (002)-reciprocal lattice reflection of the 320 nm thick c-GaN layer grown on a-C covered 3C-SiC/Si(001) quasi-substrate. g) 1x1 µm² and 10x10 µm² AFM image of a thin c-AlN nucleation layer grown on an a-C covered c-AlN/3C-SiC/Si(001) quasi-substrate at 760 °C.

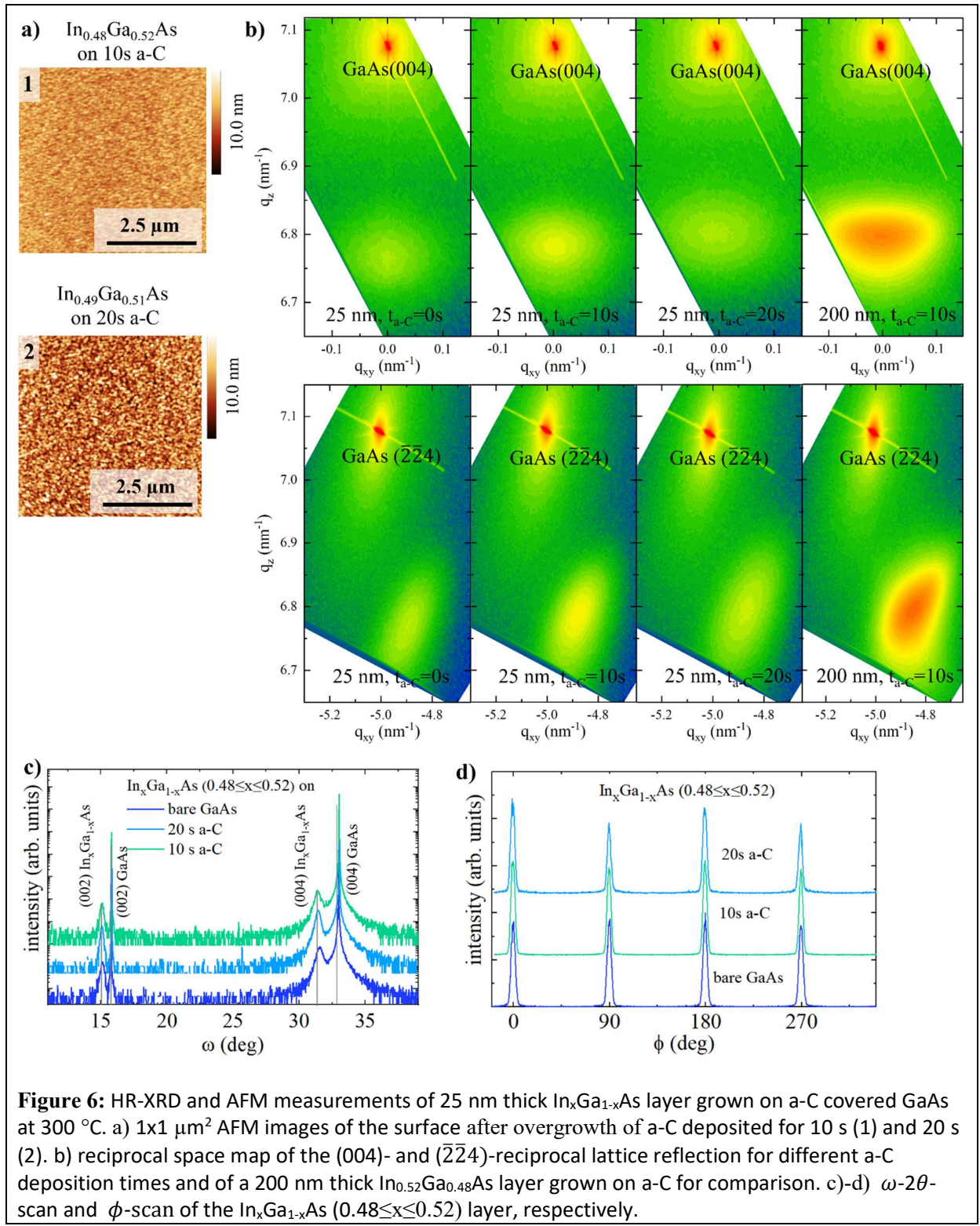

**Figure 6:** HR-XRD and AFM measurements of 25 nm thick $In_xGa_{1-x}As$ layer grown on a-C covered GaAs at 300 °C. a) 1x1 µm² AFM images of the surface after overgrowth of a-C deposited for 10 s (1) and 20 s (2). b) reciprocal space map of the (004)- and ($\bar{2}\bar{2}4$)-reciprocal lattice reflection for different a-C deposition times and of a 200 nm thick $In_{0.52}Ga_{0.48}As$ layer grown on a-C for comparison. c)-d) $\omega$-$2\theta$-scan and $\phi$-scan of the $In_xGa_{1-x}As$ (0.48≤x≤0.52) layer, respectively.

|  | $t_{a-C}$ (sec) | indium fraction (%) | (004)-reflection FWHM (deg) | relaxation (%) | dislocation density ($\times 10^{10}$ cm$^{-2}$) |
|---|---|---|---|---|---|
| 25 nm $In_xGa_{1-x}As$ on a-C | 0 | 48±1 | 0.80±0.05 | 65±1 | 1.35±0.08 |
|  | 10 | 48±1 | 0.80±0.05 | 75±1 | 1.35±0.08 |
|  | 20 | 49±1 | 1.30±0.05 | 83±1 | 3.58±0.14 |
| 200 nm $In_xGa_{1-x}As$ on a-C | 0 | 50±1 | 0.70±0.05 | 79±1 | 1.04±0.07 |
|  | 10 | 52±1 | 0.75±0.05 | 90±1 | 1.19±0.08 |

**Table 1:** Results of HR-XRD measurements of $In_xGa_{1-x}As$ layers grown on a-C covered GaAs substrates at 300 °C. The dislocation density is calculated employing the procedure outlined in [32].

**Methods**

**Fabrication of a-C layers**

Thin a-C layers were prepared on quarters of 3" GaAs and InP wafers and on 10x10 mm$^2$ pieces cleaved from 3" 3C-SiC/Si(001) wafers. The native oxide layer on the substrates was removed by a gentle hydrogen-argon plasma process employing a chamber pressure of 40 mTorr, 9 W RF-power and 10 sccm/5 sccm hydrogen/argon flux following subsequent amorphous carbon deposition employing a methane-argon plasma with a chamber pressure of 50 mTorr, 0 W-10 W RF-power, 300 W ICP-power and 35 sccm/5 sccm methane/argon flux. Both processes were conducted in an *Oxford Instrument Plasma Plus80* system.

**Sample characterization**

Thin a-C layers were analyzed by XRR measurements, Raman- and XPS-spectroscopy and AFM measurements. The crystal structure and quality of the epitaxial layers were analyzed by TEM and HR-XRD and the surface morphology by AFM measurements. HR-XRD and XRR measurements were performed employing a *Rigaku Smart-Lab* system operating at 45 kV and 200 mA, Raman-spectra by a *Renishaw inVia Raman microscope* with excitation at 532 nm and x20 magnification and AFM measurement by a *Bruker Dimension Icon* instrument. XPS was performed by means of an Omicron ESCA+ system at a base pressure of <1×10$^{-8}$ Pa. An XM1000 source with monochromatic Al Kα-radiation (1486.7 eV) was used. The angle between source and analyzer was 102 °. The take-off angle with respect to the surface plane was 60 °. Measurements were performed without neutralization at a constant pass energy of 100 eV for survey and 20 eV for element spectra. Data evaluation was performed using CasaXPS Version 2.3.23PR1.0. Atomic level structural studies were performed using ARM200F microscope with probe and image aberration CEOS correctors operating at 200 kV. Annular dark field images were obtained with a JEOL annular field detector with an inner angle of 70 mrad, a fine imaging probe of ~23 pA and convergence semiangle of ~22 mrad.

**Epitaxial growth**

For overgrowth, the carbon coated GaAs and InP samples were introduced into a III-V solid source MBE system (*Dr. Eberl MBE Komponenten GmbH*), where they were first baked 1 h at 200 °C in ultra-high vacuum and then transferred into the growth chamber. The samples were annealed for 5 min at 615 °C under arsenic (As) overpressure to desorb any adsorbents, especially attached hydrogen atoms from the plasma deposition process. Then, the sample temperature was ramped to growth temperature and the As-valve was closed. The substrate temperature was measured by a kSA-BandiT via band-edge absorption. We initialized the growth by a monolayer of group III elements as recommended in [6,7] following subsequent overgrowth to form a nucleation layer. The growth rate was 1 Ås$^{-1}$ and the As to metal ratio ≈30 with an As-flux of $p_{As}=1.5\times10^{-5}$ mbar.

Growth of c-AlN buffer layers on bare 3C-SiC/Si(001) substrates and overgrowth of c-AlN and c-GaN on carbon covered a-AlN/3C-SiC/Si(001) substrates was performed in a *Riber32* plasma assisted molecular beam (PAMBE) system. The 3C-SiC/Si(001) substrates were cleaned prior to overgrowth by ten gallium and aluminum flashes at 760 °C and 890 °C, respectively, resulting in an emergence of (2x2) surface reconstruction lines [35,38]; a-C covered samples were not treated by this procedure but were cleaned after transfer (through air) from the MBE to the PECVD employing the aforementioned hydrogen-argon plasma step. The c-AlN growth was initialized by ramping the substrate temperature to 750 °C and depositing Al for 10 s following 10 s break to cover the sample surface with a monolayer Al. Then, Al and N are deposited for 10 cycles of 30 s deposition following 30 s break at 750 °C. The Al beam equivalent pressure was $p=4.1\times10^{-5}$ mbar. c-GaN was grown at 845 °C at a Ga beam equivalent pressure of $p=1.6\times10^{-6}$ mbar. For c-AlN and c-GaN the N-plasma cell was operated at 260 W RF-power and 0.5 sccm $N_2$ flux.

**Epitaxial lift-off**

Epitaxial lift-off was performed following the approach presented in [7] in our home-built thermal evaporation chamber. First a 30 nm thin titanium adhesion layer is deposited followed by roughly 760 nm nickel deposition from alumina coated thermal evaporation boats. The chamber pressure was $<10^{-5}$ mbar. Thermal release tape was carefully stuck to the stack to peel of the layer from the substrate.

**Supplementary Information**

**Supplementary Note 1: Thickness of a-C layers measured by AFM and XRR**

We measured the thickness variation of the deposited a-C layer by performing several XRR-scans at different position on ¼ of 3" wafers; see Supplementary Fig. 1 a). Qualitatively, the individual measured spectra are very similar for the different measured positions, i.e. the maxima and minima of the Kiessig fringes exhibit the similar diffraction angles. We determined the amorphous carbon thickness at each position individually by fitting the observed Kiessig fringes and added the resulting a-C layer thicknesses to the graph. Over a large area of the wafer of roughly 20x20 mm, the amorphous carbon thickness is homogenous with <1% deviation (we usually centre the wafer in our PECVD system with respect to the PECVD table. Towards the edge of the ¼ wafer the thickness decreases by roughly 10% which we attribute to the PECVD machine table size limitations and the fact that this part of the substrate is most far away from the centre of the PECVD table and therefore is exposed to slightly different plasma conditions.

We exemplary verified the thickness of the deposited a-C layer by employing a shadow mask approach to partially cover the GaAs substrate during a-C deposition. A piece of a Si wafer was put on the GaAs to cover a part of the GaAs substrate. Then a-C was deposited for 5 min employing the method provided in the manuscript following take-off of the Si piece. Supplementary Fig. 1 shows b) an optical image, c) an 90x90µm$^2$ AFM image and d) the derived line profile of the edge between bare GaAs and a-C covered GaAs. The a-C layer thickness for this sample was derived by XRR to $\approx (13 \pm 1.5)$ nm as discussed above. A quadratic background was fitted to the line profile in d) and subtracted from the measured raw data. The so corrected height profile reveals a step of around 9-10 nm height. Please note that we assume that the roughly 2 nm thick native oxide layer on the bare GaAs is etched only away by the plasma on the uncovered GaAs areas. Taking this into account the edge height must be corrected by adding 2-3 nm resulting in a total height of 11-12 nm. These results show that the a-C layer is indeed deposited onto the GaAs substrate and that the a-C does not diffuse into the GaAs forming an GaAs-C interlayer.

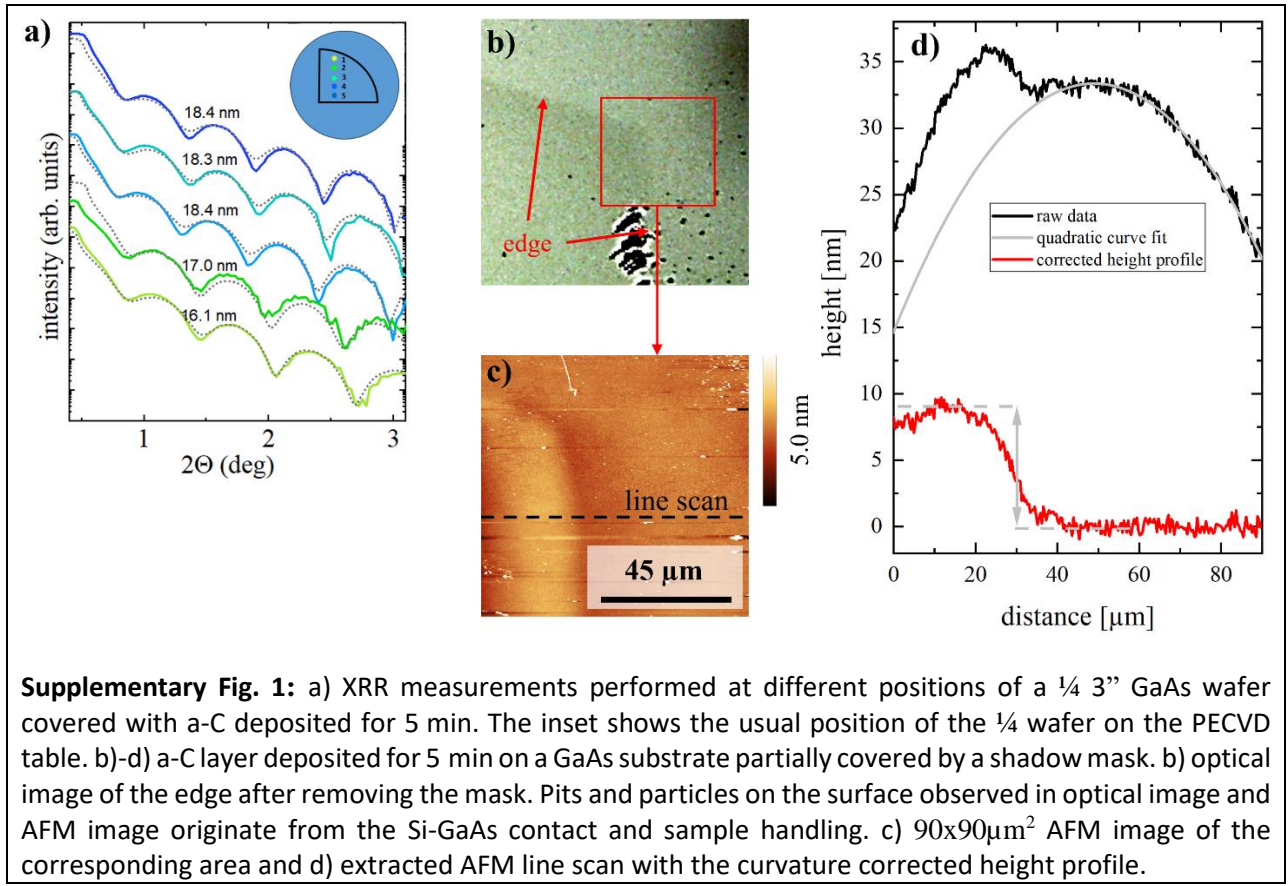

**Supplementary Fig. 1:** a) XRR measurements performed at different positions of a ¼ 3" GaAs wafer covered with a-C deposited for 5 min. The inset shows the usual position of the ¼ wafer on the PECVD table. b)-d) a-C layer deposited for 5 min on a GaAs substrate partially covered by a shadow mask. b) optical image of the edge after removing the mask. Pits and particles on the surface observed in optical image and AFM image originate from the Si-GaAs contact and sample handling. c) 90x90μm² AFM image of the corresponding area and d) extracted AFM line scan with the curvature corrected height profile.

**Supplementary Note 2: Growth of partially carbon coated surfaces**

Heating our GaAs substrate to 600° C for 5 min under As flux leads to the typical formation of pits in the surface (so called deoxidation pits). GaAs substrate surfaces covered partially with a-C show reduced surface degradation. We were still able to observe pits in AFM even when we overgrow them by 25 nm GaAs at 300 °C and III-V ratio of $\approx$ 30 making the pits a proper indicator of a-C surface coverage. Although overgrowth is not necessary to observe the pits, e.g. in Fig. 1c) we also show that we observe the pits without overgrowth, it allows us to identify if growth was performed on a closed a-C layer or on a partially a-C covered substrate ex-situ. This simple test allows to easily check for amorphous carbon deposition process reliability. Supplementary Fig. 2 b)-d) shows the impact of the a-C deposition time to the wafer surface morphology. The larger the a-C deposition time, the lower is the pit density.

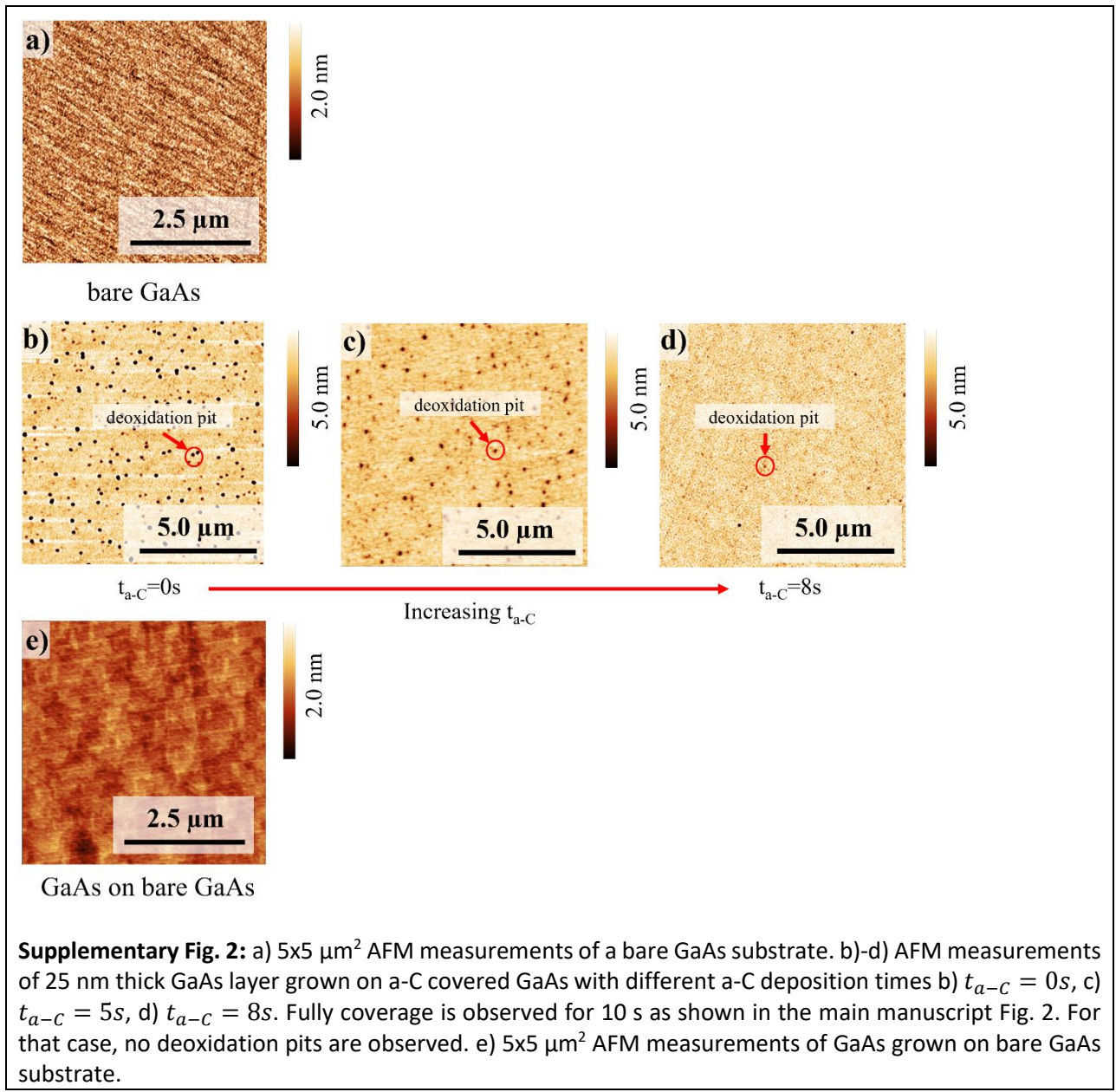

**Supplementary Fig. 2:** a) 5x5 µm² AFM measurements of a bare GaAs substrate. b)-d) AFM measurements of 25 nm thick GaAs layer grown on a-C covered GaAs with different a-C deposition times b) $t_{a-C} = 0s$, c) $t_{a-C} = 5s$, d) $t_{a-C} = 8s$. Fully coverage is observed for 10 s as shown in the main manuscript Fig. 2. For that case, no deoxidation pits are observed. e) 5x5 µm² AFM measurements of GaAs grown on bare GaAs substrate.

**Supplementary Note 3: XPS-spectra of GaAs**

Detailed analysis of the C1s peak of the a-C covered samples together with a HOPG reference sample presented in Fig. 1 d) were performed. The XPS elemental spectra were analyzed using CasaXPS software [1]. To analyze the ratio of $sp^2$- to $sp^3$-bonded carbon atoms, we followed the approach outlined in [2,3]. A Highly Oriented Pyrolytic Graphite (HOPG) reference sample was used to identify the characteristic peak shape of $sp^2$-hybridised carbon. This identification was further confirmed by analyzing a substrate coated with a single layer of graphene from [4]. We then applied the same approach to a-C covered GaAs samples ($t_{a-C}$=10 s and $t_{a-C}$=300 s). For an a-C deposition time of $t_{a-C}$=10 s, a clear $sp^2$-peak shape is observed (see Fig. 1 d)), indicating a dominant presence of $sp^2$-hybridized carbon. Additionally, a symmetric peak is observed in the O-C=O region, indicating the presence of carbonyl groups. For an a-C deposition time of $t_{a-C}$=300 s a more symmetric peak compared to the HOPG reference sample is observed (see Fig. 1 d)). To account for this, we deconvoluted this peak, revealing an additional, symmetrical component corresponding to $sp^3$-hybridized carbon alongside the $sp^2$-hybridized carbon component. The exact peak position and peak width are unknown. Thus, we applied a range of values for the $sp^3$- to $sp^2$-energy shift (based on Table 1 in [2]). This results in a composition interval as shown in Supplementary Table 1, rather than a single fixed value. Finally, we expect that $sp^2$-hybridization of the a-C is more prevalent than $sp^3$-hybridization, based on our initial assumptions.

| Decomposition of C1s | $sp^2$ LA(0.86,2.3,108) | | | Sat GL(30) | $sp^3$ GL(30) | | | O-C=O GL(30) | | |
|---|---|---|---|---|---|---|---|---|---|---|
| | Pos. (eV) | FWHM | % | % | Pos (eV) | FWHM | % | Pos (eV) | FWHM | % |
| HOPG | 284.2 | 0.69 | 93.0 | 7.0 | - | - | - | - | - | - |
| GaAs +Graphen | 284.3 | 0.98 | 97.3 | 2.7 | - | - | - | | - | - |
| GaAs +a-C (10 sec) | 284.6 | 1.27 | 91.3 | 6.5 | - | - | - | 289.4 | 1.1 | 2.2 |
| GaAs +a-C (300 sec) | 284.30-284.34 | 0.98-1.11 | 64.8-78.8 | 3.5-4.0 | 284.6-284.8 | 0.81-0.99 | 31.2-18.4 | - | - | - |

**Supplementary Table 1:** Decomposition of C1s peak for a-C covered GaAs substrates together with HOPG and GaAs+graphene reference sample. The RSF-values are 1.0 for all peaks.

We further derive the atomic ratios of the observed elements in our a-C coated GaAs templates (see Supplementary Table 2). The main contributions are carbon, Ga and As. The oxygen fraction is low (<12%) for all samples.

| | C 1s (at%) | O 1s (at%) | Ga 3d (at%) | As 3d (at%) |
|---|---|---|---|---|
| HOPG | 99.6 | 0.4 | - | - |
| GaAs +graphene | 45.0 | 9.1 | 26.0 | 19.9 |
| GaAs +a-C (10 sec) | 25.8 | 11.8 | 31.1 | 31.3 |
| GaAs +a-C (300 sec) | 78.5 | 4.7 | 10.0 | 6.8 |
| | | | | |

**Supplementary Table 2:** Atomic ratios of the a-C covered GaAs substrates together with HOPG and GaAs+graphene reference samples. The RSF-values are 1.0 for all peaks.

Beside the C-related XPS-spectra of the a-C layer presented in the main manuscript, we also investigated features originating from the a-C covered GaAs substrate. During transport to the XPS system the samples were exposed to air which could in principle causes oxidation of the GaAs substrate under the a-C layer. Supplementary Fig. 3 shows the XPS-spectra of a GaAs substrate covered by a-C deposited for 10 s and 300 s respectively. Weak $As_2O_3$ and $Ga_2O_3$ related peaks are detected in these spectra. The $As_2O_3$ peak is slightly larger for 10 s compared to 300 s a-C deposition which could be attributed to the monolayer-like thickness of the a-C layer and thus a thinner protection barrier against substrate oxidation or $As_2O_3$ and $Ga_2O_3$ oxides are not completely removed by the H-plasma cleaning step. However, the signal intensity is quite weak compared to values obtained for dry transferred graphene-GaAs surfaces [5]. For comparison, we added an XPS-spectrum of a graphene covered GaAs substrate; the oxide was not removed from this GaAs substrate. A significantly larger oxide related $As_2O_3$ and $Ga_2O_3$ is observed for this sample. The results demonstrate that the a-C protects the GaAs surface from significant oxidation.

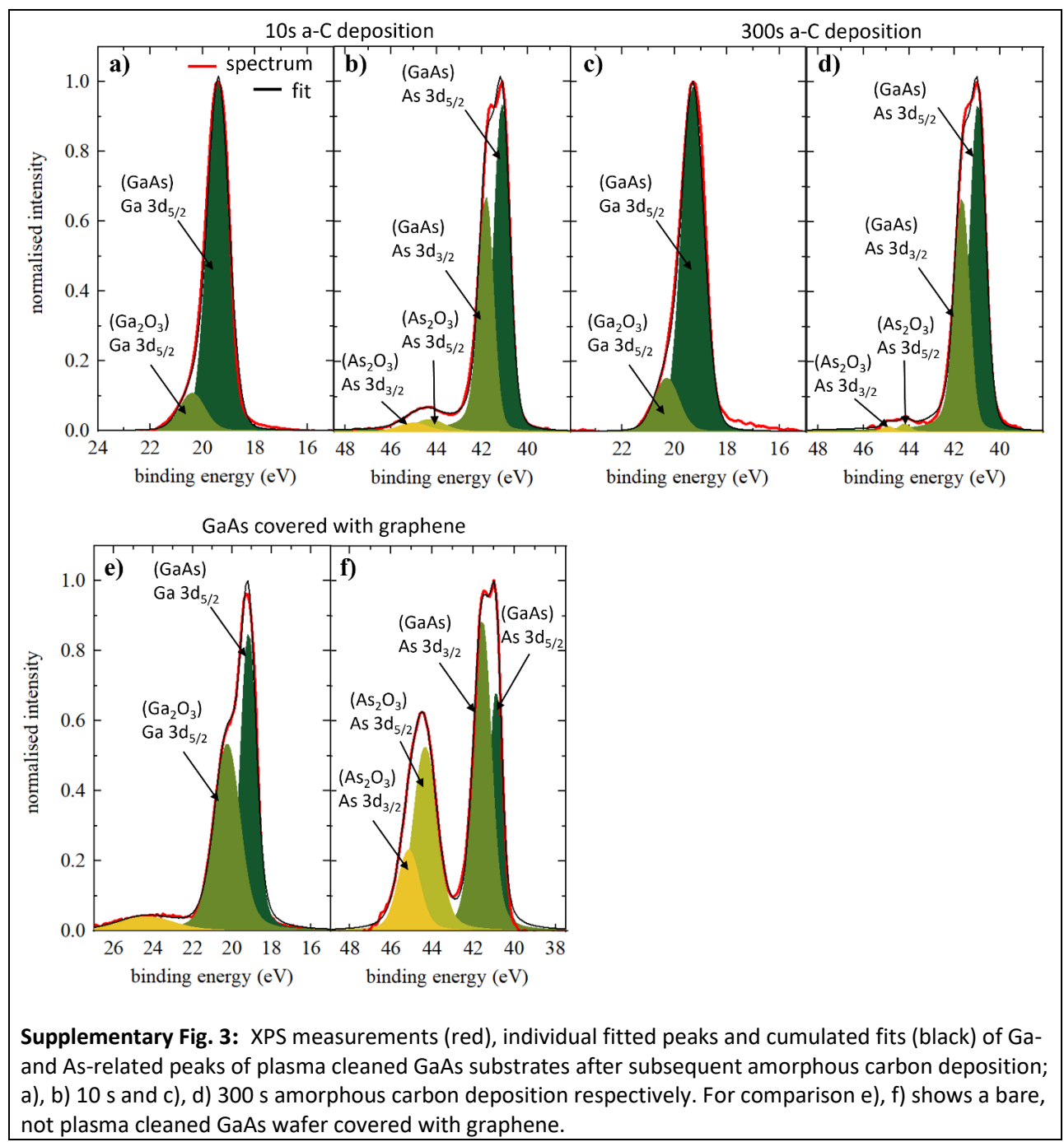

**Supplementary Fig. 3:** XPS measurements (red), individual fitted peaks and cumulated fits (black) of Ga- and As-related peaks of plasma cleaned GaAs substrates after subsequent amorphous carbon deposition; a), b) 10 s and c), d) 300 s amorphous carbon deposition respectively. For comparison e), f) shows a bare, not plasma cleaned GaAs wafer covered with graphene.

**Supplementary Note 4: Dislocation density calculation**

We performed $\omega$-scans of the layers (004)-reciprocal lattice reflection and derived the FWHM of the peaks. The dislocation density is estimated using the following equation with the dislocation Burger's vector modulus of 4 Å as derived in [6]:

$$n = \frac{FWHM^2}{9b^2}$$

We observe equal dislocation densities from measurements including the [110] and [$\bar{1}$00] direction, i.e., in-plane sample rotation by 90 ° as discussed in [6]. Within the GaAs/In$_x$Ga$_{1-x}$As material system, layer tilt is well known and is attributed to the different crystal slip planes e.g., the (111) and ($\bar{1}\bar{1}\bar{1}$) plane so the relaxation of a mismatched layer is not homogeneous. A dominant relaxation in [110] direction occurs for growth on bare GaAs substrates but not for growth on a-C.

In comparison to our HR-XRD measurements we performed TEM measurements (see Supplementary Fig. 4) of a 25 nm thick In$_{0.09}$Ga$_{0.91}$As layer grown on a monolayer like a-C layer (t$_{a-C}$=10 s). On larger magnification most of the layer is free of stacking faults, while several stacking faults are observed on smaller magnification. The left side of the measurement presented in Supplementary Fig. 4 b) (1) is free of dislocations, while we observe a locally larger density on the right side (2). We estimate a stacking fault density of the selected area of >10$^{10}$ cm$^{-2}$. We speculate that the stacking fault density might vary locally on nanometer scale because of thickness variation of the a-C layer. As discussed in the manuscript, the stacking fault density is very sensitive to the amorphous carbon layer thickness.

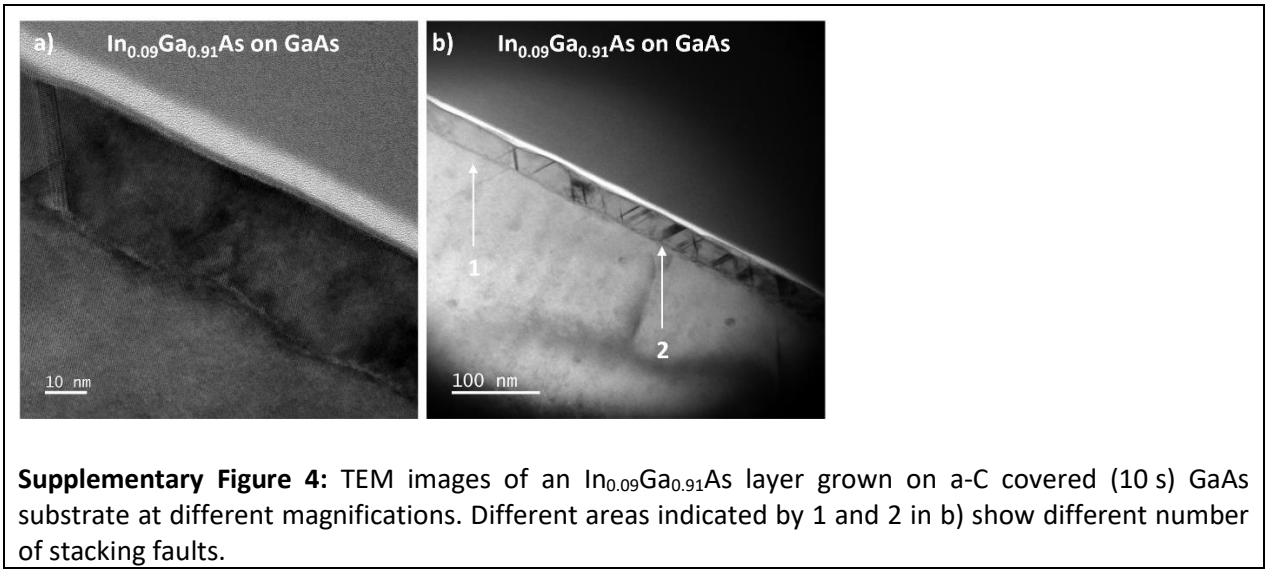

**Supplementary Figure 4:** TEM images of an In$_{0.09}$Ga$_{0.91}$As layer grown on a-C covered (10 s) GaAs substrate at different magnifications. Different areas indicated by 1 and 2 in b) show different number of stacking faults.

**Supplementary Note 5: Growth of thick layers on a-C covered Substrates**

Heterostructure and device fabrication requires growth of thick layers. We performed remote epitaxial growth of 200 nm thick GaAs and AlAs layers on a-C covered GaAs at 300 °C and III-V ratio of $\approx$ 30. Supplementary Fig. 5 a) and b) show the surface morphology of both films. Smooth surfaces are obtained with a roughness $\leq$0.4 nm on a 10x10 µm$^2$ scale demonstrating successful growth of thick layers. Epitaxial lift-off of thicker films was performed by depositing a nickel stressor on the layer surface following subsequent peeling-off the film. An exemplary image of a 300 nm $In_{0.1}Ga_{0.9}As$ film peeled from the substrate but still on the Ni-film/thermal release tape stack is shown in Supplementary Fig. 5c) together with AFM measurements in d) of the backside of the film after lift-off, $\theta/2\theta$-measurement in e) and Raman measurements in f). The $\theta/2\theta$-spectrum reveals decent peaks indicating the Ni, $TiO_2$, and $In_{0.1}Ga_{0.9}As$ layer and reveals successful $In_{0.1}Ga_{0.9}As$ layer lift-off. We speculate that the $TiO_2$ signal originates from the oxidation of the deposited Ti layer. Raman measurements of the $In_{0.1}Ga_{0.9}As$ film and GaAs substrate after lift-off show two prominent features: First, a shift of the GaAs-like LO peak from 291 $cm^{-1}$ (substrate) to 282 $cm^{-1}$ (lift-off film). Second, disorder activated TO mode in the lifted off film indicated by the peak at 257 $cm^{-1}$. Following the discussion in [7,8] we speculate that the shift in the GaAs LO peak and the arise of disorder activated TO modes might be due to strain inhomogeneities and plastic relaxation in the lifted off film introduced by the attached nickel stressor. The AFM images of the back side of the lifted-off film reveal surface undulations which probably arise from deformation of the lifted-off film induced by the Ni stressor which is still attached to the layer.

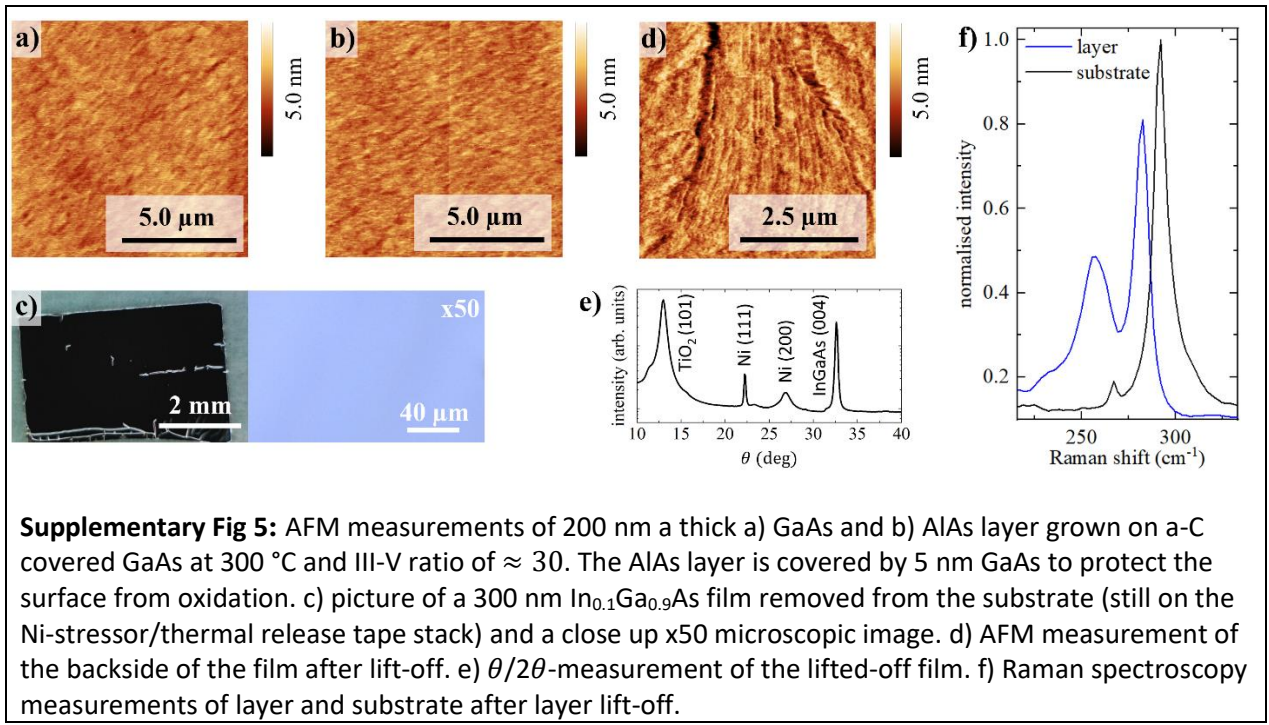

**Supplementary Fig 5:** AFM measurements of 200 nm a thick a) GaAs and b) AlAs layer grown on a-C covered GaAs at 300 °C and III-V ratio of $\approx 30$. The AlAs layer is covered by 5 nm GaAs to protect the surface from oxidation. c) picture of a 300 nm $In_{0.1}Ga_{0.9}As$ film removed from the substrate (still on the Ni-stressor/thermal release tape stack) and a close up x50 microscopic image. d) AFM measurement of the backside of the film after lift-off. e) $\theta/2\theta$-measurement of the lifted-off film. f) Raman spectroscopy measurements of layer and substrate after layer lift-off.

**Supplementary Note 6: Fabrication of optically active structure**

We fabricated droplet etched $In_{0.57}Ga_{0.43}As$ quantum dots embedded in an $In_{0.52}Al_{0.48}As$ matrix grown on an a-C covered InP substrate (for more details see [9]) to show that our a-C covered templates allow the fabrication of optically active heterostructures. We performed photoluminescence measurements at 16 K and 4.5 mW excitation power at 635 nm (see Supplementary Fig. 6). A broad peak is detected clearly in the spectrum revealing successful growth of an optically active structure on our a-C covered InP substrates. We attribute the broad peak shape to the non-optimised growth parameters on the a-C covered InP substrate, i.e. the quantum dot size distribution might be large.

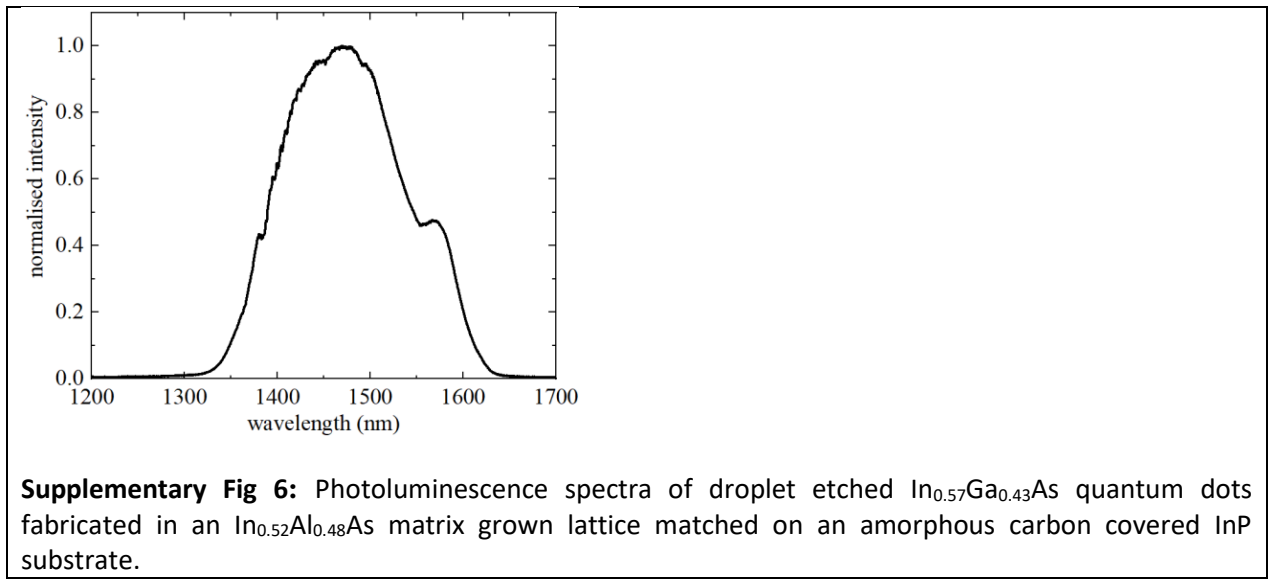

**Supplementary Fig 6:** Photoluminescence spectra of droplet etched $In_{0.57}Ga_{0.43}As$ quantum dots fabricated in an $In_{0.52}Al_{0.48}As$ matrix grown lattice matched on an amorphous carbon covered InP substrate.